\documentclass[article,onecolumn,amsmath,amssymb,aps,
]{revtex4}

\usepackage{epsfig,bm,epsf,graphics}
\usepackage{graphicx}% Include figure files
\usepackage{dcolumn}% Align table columns on decimal point
\usepackage{bm}% bold math
\usepackage{xcolor}

\begin{document}
%\draft
\preprint{APS/123-QED}

\title{Frustrated Spin Systems: History of the Emergence of a Modern Physics }% Force line breaks with \\

\author{ Hung T. Diep\footnote{diep@cyu.fr, corresponding author}}
\affiliation{%
 Laboratoire de Physique Th\'eorique et Mod\'elisation,
CY Cergy Paris Universit\'e, CNRS, UMR 8089\\
2, Avenue Adolphe Chauvin, 95302 Cergy-Pontoise Cedex, France.\\
}% }%



\date{\today}% It is always \today, today,


\begin{abstract}

In 1977, G\'erard Toulouse has proposed a  new concept termed as "frustration" in spin systems. Using this definition, several frustrated models have been created and studied, among them we can mention the Villain's model, the fully frustrated simple cubic lattice, the antiferromagnetic triangular lattice. The former models are systems with mixed ferromagnetic and antiferromagnetic bonds, while in the latter containing only an antiferromagnetic interaction, the frustration is caused by the lattice geometry.  These frustrated spin systems have novel properties that we will review in this paper. One of the striking aspects is the fact that well-established  methods such as the renormalization group fail to deal with the nature of the phase transition in frustrated systems. Investigations of properties of frustrated spin systems have been intensive since the 80's.   I myself got involved in several investigations of frustrated spin systems soon after my PhD. I have learned a lot from  numerous discussions with G\'erard Toulouse. Until today, I am still working on frustrated systems such as skyrmions. In this review, I trace back a number of my works over the years on frustrated spin systems going from exactly solved 2D Ising frustrated models, to XY and Heisenberg  2D and 3D frustrated lattices.  At the end I present my latest results on skyrmions resulting from the frustration caused by the competition between the exchange interaction and the Dzyaloshinskii-Moriya interaction under an applied magnetic field. A quantum spin-wave theory using the Green's function method is shown and discussed.
\vspace{0.5cm}
\begin{description}
\item[PACS numbers: 5.10.Ln;64.30.+t;75.50.Cc]
\end{description}
\end{abstract}

%\pacs{PACS numbers: XXX}% PACS, the Physics and Astronomy
\keywords{Frustrated Spin Systems, Exactly Solved Models, Non-collinear Spin Configuration, Phase Transitions, Reentrance, Disorder Lines}

\maketitle

%\tableofcontents

\section{Introduction}
This paper is devoted to the memory of G\'erard Toulouse. Since 1981, I was interested in the effects of frustation in spin systems, a few years after G. Toulouse \cite{Toulouse} and J. Villain \cite{Villain} independently introduced the notion of frustration. Note that  in the early years of the 70's, the introduction of the renormalization group theory \cite{Wilson,Amit,Cardy,Zinn} and the exactly solved spin systems \cite{Baxter} have made a great progress in the understanding of the mechanism governing the phase transition: one can mention the distinction between first- and second-order phase transitions, the discovery of universality classes distinguished by critical exponents and the theory of finite-size scaling.  Competing interactions have been studied in many investigations by these advanced methods but most of these studies are restrained to the cases where the spin ordering is collinear, namely ferromagnetic or antiferromagnetic (exceptions made for $q$-state Potts models \cite{Baxter}).  The first frustrated spin system with non-collinear spin  ordering is the helimagnetic structure discovered independently by Yoshimori \cite{Yoshimori} and  Villain \cite{Villain59}: the competition between the ferromagnetic interaction between nearest neighbors (NN) $J_1>0$ and the antiferromagnetic  interaction $J_2<0$ between next NN (NNN) results in a non-collinear spin configuration when $|J_2|/J_1$ is larger than a critical value.  We return to this case below. The most popular case which has been widely studied since the 80's is the antiferromagnetic triangular lattice with vector spins (XY and Heisenberg spins). Note that the antiferromagnetic triangular lattice with Ising spins has been exactly solved in 1950 by Wannier \cite{Wannier1,Wannier2}.  As will be recalled below,  the ground state (GS) of the antiferromagnetic triangular lattice with vector spins is the well-known $120^{\circ}$ structure. The  nature of the phase transition in the case  of stacked triangular lattices with vector spins has been subject of investigations for more than 30 years
 \cite{Kawamura1988,Kawamura1998,Delamotte,Delamotte2010,Reehorst,Delamotte2024,NgoDiepHeis,NgoDiepXY}. We will recall this case in this review.  In the 80's I have studied alone or with collaborators a number of frustrated systems such as the fully frustrated simple cubic lattice \cite{FFSC1,FFSC2,FFSC3} (G. Toulouse was a coauthor in \cite{FFSC1}).  Note that this  system has been studied a few years before by Derrida et al. \cite{Derrida1,Derrida2}. I also studied during this period  the  helimagnets (spin waves and phase transition), the phase transition in the frustrated face-centered cubic (FCC) Heisenberg lattice and the frustrated Villain's model with XY spin model.  Then I started with my colleagues a series of papers on exactly solved Ising frustrated models such as the generalized  Kagom\'e lattices, honeycomb lattice and dilute centered square lattices.  We found new striking phenomena such as multiple phase transitions,  reentrance phases,  disorder lines, and coexistence of order and disorder.  I will show some of these results below.  Going further into the frustration  effects, I studied with colleagues the effect of Dzyaloshinskii-Moriya (DM) interaction in ferromagnetic and antiferromagnetic materials. As a result of the competition between collinear ordering and perpendicular spin ordering due to the DM interaction, non-collinear orderings are observed. Under a perpendicular applied magnetic field, the spin configuration can turn into skyrmions arranged on an ordered way which is called "skyrmion crystal". A skyrmion is a topologically stable vortex-like structure which contains a dozen of spins turning around a center. I will recall some of the results below.

Important works on many aspects of frustrated systems have been reviewed  in a book \cite{DiepFSS}. The reader is referred to this reference for a rather complete bibliography. 

This paper is a personal account of works on a subject inspired by the work of G. Toulouse on the frustration.
In section \ref{Frustration}, I recall some elementary notions of the frustration and its effect on the GS. In section \ref{Exact}, I review some results on exactly solved Ising frustrated systems. In section \ref{FFSystems}, I present some results of my early works on frustrated systems either in collaboration with G. Toulouse  or inspired by discussions with him. I also recall here results on the nature of the phase transition of other frustrated systems such as helimagnets, fully frustrated antiferromagnetic HCP and FCC lattices with vector spins. A brief account on the nature of the phase transition in the stacked antiferromagnetic triangular lattice is presented. Some basic results on skyrmions are discussed in section \ref{Skyr}.  Finally, a quantum theory of magnons using the Green's function method for non-collinear spin configurations such as the ones in helimagnets and in systems with Dzyaloshinskii-Moriya is shown in section \ref{Green}. Concluding remarks are given in section \ref{Concl}.

\section{Frustration: Introduction}\label{Frustration}
Before introducing the definition of the frustration given by Toulouse \cite{Toulouse} in the context of spin glasses, let us recall some early developments in the 60's and early 70's on spin glasses. The new quenching experimental method introduced by Pol Duwez \cite{Duwez} in 1960 has resulted in the fabrication on many metglasses (metallic glasses) in which bond disorder causes metastable states.  The reader is referred to the reference \cite{review} for reviews on works up to 1984 on metglasses. Several theoretical models of spin glasses have been published in the 70's. One can mention two of them: the Edwards-Anderson (EA) model \cite{EAmodel} and the Sherrington-Kirpatrick (SK) model \cite{SKmodel}.  Spin glasses are viewed as systems having two main ingredients: bond disorder and frustration. In a spin glass, there  is a random mixing of ferromagnetic and antiferromagnetic bonds between spins. This mixing causes the frustration as we willl define below. The EA model consists of nearest-neighbor Ising spins interacting with each other via a pairwise random interaction. This model has been exactly solved by its authors using the replica method with the hypothesis that the random interactions obey a Gaussian distribution covering negative and positive values. The  SK model was also based on the Ising spin model, however the spins interact with each other via an infinite range with a Gaussian distribution with positive and negative interactions.  The SK model has also been solved by its authors using the replica method, but  the entropy is found negative at low temperatures.  This model was finally solved by a  replica breaking ansatz  introduced by Parisi in 1979 \cite{Parisi} where stable spin glass states at low $T$ are found. The reader is referred to the book by M\'ezard, Parisi and Virasoro \cite{Mezard} for details on the replica method. 

Let us note that in the early 80's numerical simulations have also been developped to study spin glasses. I myself started to write Monte Carlo programs to study several spin glass models. I mention here a few works I published with co-authors on spin glass using Monte Carlo simulations during this period \cite{Nagai83,Diep84,Diep86}. In this period, there was the nice work of Ogielski on the $\pm J$ Ising model on the simple cubic lattice, using Monte Carlo simulations. His results are in agreement with experiments performed on insulating spin glasses,  despite the simplicity of the discrete model for spin glasses. 

As said above, the two main ingredients of spin glass are  bond disorder and frustration. The idea to separate the two ingredients to study systems with frustration but without bond disorder comes naturally: such frustrated systems having periodical $\pm$ bond distributions on the lattice are subject to exact treatment. This is what I have done in a series of papers on exactly solved models with my young team in the late 80's (see section \ref{Exact}).

We have seen that the frustration is caused by the competition between various interactions in the system. However, the frustration can also be generated by only one type of interaction in the case where the lattice structure which does not allow to fully satisfy every interaction bond: this is the case of a triangular lattice with an antiferromagnetic interaction between nearest neighbors (NN), or the face-centered cubic (FCC) lattice and the hexagonal-closed-packed (HCP) lattice, with antiferromagnetic NN interaction. These latter cases are called "geometry frustration".  

The effects of the frustration are numerous. We can mention a few of them (the reader is referred to chapter 1 of Ref. \cite{DiepFSS}): (i) the GS degeneracy is very high, in many cases the GS degeneracy is infinite, (ii) in the case of vector spins (XY and Heisenberg spins), the GS spin configuration is non-collinear, unlike the collinear configuration in ferromagnets and antiferromagnets.  A famous example is the 120-degree spin structure in the antiferromagnetic triangular lattice (note that the helimagnetic structure has been determined a long time ago \cite{Yoshimori,Villain59} (iii) the nature  of the phase transition is often difficult to determine. To our knowledge, the phase transition in most (if not all) frustrated systems in 3D is of the first-order transition. Well-known examples are the cases of HCP antiferromagnets \cite{Diep1992HCP,HoangDiep2012HCP}, the FCC antiferromagnets \cite{DiepKawamura} and the antiferromagnetic stacked triangular lattice \cite{Kawamura1988,Kawamura1998,Delamotte,Delamotte2010,Reehorst,Delamotte2024,NgoDiepHeis,NgoDiepXY}. 

Let us examine some simple cases in the following.

\subsection{Definition}\index{frustration, definition}
We give below some popular cases for readers who are not familiar with frustrated spin systems.

\begin{figure}[ht]
\centering
\includegraphics[width=3.2 in]{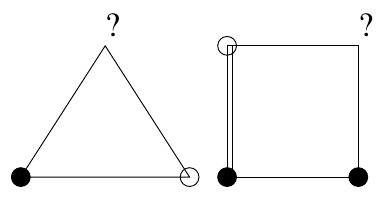}
%\begin{figure}
%\centerline{\epsfig{file=fcc06_submitted/costhetas.pdf,width=3.2in}} 
\caption{Examples
of frustrated  plaquettes:  ferro- and antiferromagnetic
interactions, $J$ and $-J$, are shown by single and double lines,
$\uparrow$ and $\downarrow$ Ising spins by black and void circles,
respectively. Choosing any orientation for the spin marked by the
question mark will leave one of its bonds unsatisfied (frustrated
bond).} \label{fig:IntroFP}
\end{figure}

In the pairwise Ising and Heisenberg model, the interaction energy between  two spins $\mathbf S_i$ and $\mathbf S_j$ is given by
 $E=-J \left(\mathbf S_i
\cdot \mathbf S_j\right)$ where $J>0$ ($J<0$) is the exchange integral giving  rise to parallel (antiparallel) spin pairs. The minimum of $E$ is $-J$ or $-|J|$.  For a crystal where the NN interaction is ferromagnetic, the GS of the system corresponds to the spin configuration where all spins
are parallel. This is true for any lattice structure. However, if $J$ is
antiferromagnetic, the spin configuration of the GS depends on the
lattice structure: i) for lattices containing no elementary
triangles, such as square lattice and simple
cubic lattices, the GS is the configuration in which each
spin is antiparallel to its neighbors, namely every bond is fully
satisfied (lowest energy); ii) for lattices containing elementary triangles such
as the triangular lattice, the FCC and HCP lattices, it is impossible to construct a GS where all bonds have the lowest energy $J$(see
Fig. \ref{fig:IntroFP}): The GS does not correspond to the minimum
of the interaction of every spin pair. In this case, one says that
the system is frustrated.

Let us recall the frustration defined by Toulouse: one considers an elementary cell of the lattice.
The lattice cell is in general a polygon
formed by faces hereafter called "plaquettes". For example, the
elementary cell of the simple cubic (SC) lattice is a cube with six
square plaquettes, the elementary cell of the FCC lattice is a
tetrahedron formed by four triangular plaquettes. Let $J_{i,j}$ be
the interaction between two NN spins of the plaquette. According
to the definition of Toulouse \cite{Toulouse}, the plaquette is
frustrated if the parameter $P$ defined below is negative
\begin{equation}
P=\prod_{\left<i,j\right>}\mathrm{sign}(J_{i,j}), \label{frust1}
\end{equation}
where the product is performed over all $J_{i,j}$ around the
plaquette. Two examples of frustrated plaquettes are shown in Fig.
\ref{fig:IntroFP}: a triangle with three antiferromagnetic bonds
and a square with three ferromagnetic bonds and one
antiferromagnetic bond.  $P$ is negative in both cases.  One sees
that if one tries to put Ising spins on those plaquettes, at least
one of the bonds around the  plaquette will not be satisfied. For
vector spins, we show below that in the lowest energy state, each
bond is only partially satisfied.

%MODIF
We consider another spin system, which can be called  frustrated system because each bond is only partially satisfied:
 this is the case with different kinds of conflicting interactions and the GS
does not correspond to the minimum of each kind of interaction.
Let us examine a chain of spins where the NN interaction
$J_1$ is ferromagnetic while the next NN (NNN) interaction $J_2$
is antiferromagnetic. As long as $|J_2|\ll J_1$, the GS is
ferromagnetic: every NN bond is then satisfied but the NNN ones
are not. Of course, when $|J_2|$ exceeds a critical value, the
ferromagnetic GS is no longer valid (see the helimagnet example below): both
the NN and NNN bonds are not fully satisfied.

%In a general manner, we can say that a spin system is frustrated
%when one cannot find a configuration of spins to fully satisfy the
%interaction (bond) between every pair of spins. In other words,
%the minimum of the total energy does not correspond to the minimum
%of each bond. This situation arises when there is a competition
%between different kinds of interactions acting on a spin by its
%neighbors or when the lattice geometry does not allow to satisfy
%all the bonds simultaneously. 

%Fig1

Let us touch upon the GS degeneracy. One notes that for the triangular plaquette, the degeneracy is
three, and for the square plaquette it is four, in addition to the
degeneracy associated with global spin reversal.  Therefore, the
degeneracy of an infinite lattice composed of such plaquettes is
infinite, in contrast to the unfrustrated case.

At this stage, it is worth noting that in the above discussion we
have assumed the interaction between two spins to be of the form
$E=-J \left(\mathbf S_i \cdot \mathbf S_j\right)$. However,  the 
frustration takes place also in systems with other types of interaction such as
the Dzyaloshinski-Moriya interaction $E=- \mathbf D \cdot \left(\mathbf S_i
\wedge \mathbf S_j\right)$: a spin system is frustrated
whenever the minimum of the system energy does not correspond to
the minimum of all local interactions, whatever the form of
interaction. In such a case, the definition of frustration
is more general than the one using Eq. (\ref{frust1}).

The determination of the GS of various  frustrated Ising spin
systems such as the Kagom\'e lattice, the honeycomb lattice and the Union-Jack lattice, 
as well as discussions on their thermodynamic properties using exactly solved methods have been presented
 in Ref. \cite{DiepGiacomini}.
In the following section, we recall the GS of a few frustrated systems with XY and Heisenberg
spins.

\subsection{Non-collinear spin configurations}\index{non-collinear spin configuration}

Consider again the plaquettes shown in Fig. \ref{fig:IntroFP}.
In the case of $XY$ spins, one can calculate the GS configuration
by minimizing the energy of the plaquette $E$ while keeping the
spin modulus constant. In the case of the triangular plaquette,
suppose that spin $\mathbf S_i$ $(i=1,2,3)$ of amplitude $S$ makes
an angle $\theta_i$ with the $Ox$ axis. Writing $E$ and
minimizing it with respect to the angles $\theta_i$, one has
\begin{eqnarray}
E&=&J(\mathbf S_1\cdot \mathbf S_2+\mathbf S_2\cdot \mathbf S_3+\mathbf S_3\cdot \mathbf S_1)\nonumber \\
&=&JS^2\left[\cos (\theta_1-\theta_2)+\cos (\theta_2-\theta_3)+
\cos (\theta_3-\theta_1)\right],\nonumber \\
\frac{\partial E}{\partial \theta_1}&=&-JS^2\left[\sin
(\theta_1-\theta_2)-
\sin (\theta_3-\theta_1)\right]=0, \nonumber \\
\frac{\partial E}{\partial \theta_2}&=&-JS^2\left[\sin
(\theta_2-\theta_3)-
\sin (\theta_1-\theta_2)\right]=0, \nonumber \\
\frac{\partial E}{\partial \theta_3}&=&-JS^2\left[\sin
(\theta_3-\theta_1)- \sin (\theta_2-\theta_3)\right]=0. \nonumber
\end{eqnarray}

A solution of the last three equations is
$\theta_1-\theta_2=\theta_2 -\theta_3=\theta_3-\theta_1 =2\pi/3$.
This solution can be also obtained by writing the following equality
$$
E=J(\mathbf S_1\cdot \mathbf S_2+\mathbf S_2\cdot \mathbf
S_3+\mathbf S_3\cdot \mathbf S_1)
=-\frac{3}{2}JS^2+\frac{J}{2}(\mathbf S_1 + \mathbf S_2 + \mathbf
S_3)^2.
$$
The minimum of the above equation evidently corresponds to $\mathbf S_1 + \mathbf
S_2 + \mathbf S_3=0$ which yields, using a geometric construction, the $120^\circ$
structure.\index{$120^\circ$ structure} This is true also for
Heisenberg spins.

We can do the same calculation for the case of the frustrated
square plaquette \cite{Berge}. Suppose that the antiferromagnetic bond $-\eta J$ connects
the spins $\mathbf{S}_1$ and $\mathbf{S}_2$ and the ferromagnetic bond is $J$ ($J>0)$. We find
\begin{equation}
\theta_2-\theta_1=\theta_3 -\theta_2=\theta_4-\theta_3
=\frac{\pi}{4} \textrm { and } \theta_4-\theta_1=\frac{3\pi}{4}
\label{frust2}
\end{equation}

If the antiferromagnetic bond is equal to $-\eta J$, the solution
for the angles is\cite{Berge}
\begin{equation}
\cos\theta_{32}=\cos\theta_{43}=\cos\theta_{14}\equiv \cos \theta=\frac
{1}{2}[\frac {\eta+1}{\eta}]^{1/2}\label{frust2a}
\end{equation}
and $|\theta_{21}|=3|\theta|$, where $\cos\theta_{ij}\equiv
\cos(\theta_{i}-\theta_{j})$.

This solution exists if $| \cos\theta |\leq 1$, namely
$\eta>\eta_c=1/3$. One can check that when $\eta=1$, one has
$\theta =\pi/4$, $\theta_{21}=3\pi/4$.

\begin{figure}[ht]
\centering
\includegraphics[width=3.2 in]{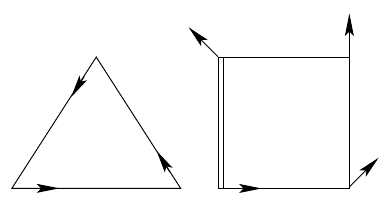}
%\begin{figure}[htb!] %Fig2
%\centerline{\epsfig{file=IntroNCFT.pdf,width=2.0in}} 
\caption{ Non-collinear spin configuration of frustrated triangular and square
plaquettes with $XY$ spins: ferro- and antiferromagnetic
interactions
 $J$ and $-J$ are indicated by thin and double lines, respectively.}
 \label{fig:IntroNCFT}
\end{figure}

We show the frustrated triangular and square lattices in Fig.
\ref{fig:IntroNCFT} with $XY$ spins ($N=2$).

%Fig2

One observes that there is  a two-fold degeneracy  resulting from
the symmetry by mirror reflecting with respect to an axis, for
example the $y$ axis in Fig. \ref{fig:IntroNCFT}. Therefore the
symmetry of these plaquettes is of Ising type O(1), in addition to
the symmetry SO(2) due to the invariance by global rotation of the
spins in the  plane.

\begin{figure}[ht]
\centering
\includegraphics[width=3.2 in]{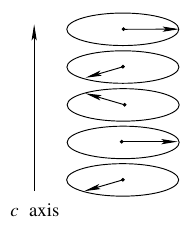}
%\begin{figure}[htb!] %Fig3
%\centerline{\epsfig{file=IntroHC.eps,width=2in}} 
\caption{Helical
configuration when $\varepsilon = |J_2|/J_1>\varepsilon_c= 1/4$
($J_1>0$, $J_2<0$).}
 \label{fig:IntroHC}
\end{figure}

Another example is the case of a chain of Heisenberg spins with
ferromagnetic interaction $J_1(>0)$ between NN and
antiferromagnetic interaction
 $J_2 (<0)$ between NNN. When
$\varepsilon = |J_2|/J_1$ is larger than a critical value
$\varepsilon_c$, the spin configuration of the GS becomes non-collinear. One shows that the helical configuration displayed in
Fig. \ref{fig:IntroHC} is obtained by minimizing the interaction
energy (see \cite{Harada,Rastelli,Diep89}):
\begin{eqnarray}
E&=&-J_1\sum_i\mathbf S_i\cdot\mathbf S_{i+1}
+|J_2|\sum_i\mathbf S_i\cdot\mathbf S_{i+2}\nonumber \\
&=&S^2\left[ -J_1\cos \theta+|J_2|\cos (2\theta)\right]\sum_i1\nonumber \\
\frac{\partial E}{\partial \theta}&=&S^2\left[J_1\sin \theta
-2|J_2|\sin(2\theta)\right]\sum_i1=0\nonumber \\
&=&S^2\left[J_1\sin \theta -4|J_2|\sin\theta \cos \theta
\right]\sum_i1=0,
\end{eqnarray}
where one has supposed that the angle between NN spins is
$\theta$.

The two solutions are
$$
\sin \theta=0 \longrightarrow \theta=0 \hspace{0.2cm}
\textrm{(ferromagnetic solution)}
$$
and
\begin{equation}
\cos \theta=\frac{J_1}{4|J_2|} \longrightarrow \theta= \pm \arccos
\left(\frac{J_1}{4|J_2|}\right).
\end{equation}\label{helichain}
The last solution is possible if $-1\le \cos\theta \le 1$, i.e.
$J_1/\left(4|J_2|\right)\le 1$ or $|J_2|/J_1\ge 1/4 \equiv
\varepsilon_c$.

As before, there are two degenerate configurations:
clockwise and counter-clockwise.

%Fig3

%Fig4
\begin{figure}[ht]
\centering
\includegraphics[width=3.2 in]{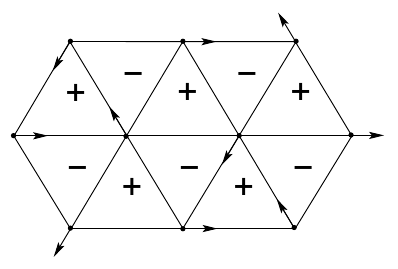}
%\begin{figure}[htb!]  %Fig4
%\centerline{\epsfig{file=IntroAFTL.eps,width=2.5in}}
\caption{\label{fig:IntroAFTL} Antiferromagnetic triangular
lattice with  $XY$ spins. The positive and negative chiralities
are indicated by $+$ and $-$.}
\end{figure}

We generate a triangular lattice using plaquettes as shown in Fig.
\ref{fig:IntroAFTL}. The GS corresponds to the state where all
triangle plaquettes of the same orientation have the same chirality:
plaquettes $\bigtriangleup$ have positive chirality
and plaquettes $\bigtriangledown$ have negative chirality.  Mapping the chiralities into the Ising spins, we have a perfect
antiferromagnetic Ising order. This order is broken at a phase
transition temperature where the chirality vanishes.

%Let us enumerate two frequently encountered frustrated spin
%systems where the NN interaction is antiferromagnetic: the FCC
%lattice and the HCP lattice. These two lattices are formed by
%stacking tetrahedra with four triangular faces.  The frustration
%due to the lattice structure such as in these cases is called
%"geometry frustration".

\section{Exactly solved frustrated models}\label{Exact}

As seen in the following, frustrated systems are very difficult to deal with because of the GS degeneracy and the GS symmetry.  They are excellent candidates to test
approximations and improve theories. Since it is hard to relate observed phenomena to real mechanisms  in real systems (disordered systems,
systems with long-range interaction, three-dimensional systems,
etc), it is worth to search for the origins of those phenomena in
exactly solved systems.  These exact results will help to
understand at least qualitatively the behavior of real systems which are in
general much more complicated.

We have exactly solved a number of 2D frustrated Ising model going from the Kagom\'e  model with NNN interaction to an anisotropic frustrated honeycomb lattice, in passing by various frustrated models \cite{Aza87,Diep91b,Diep91a,Diep92,Diep92a}. The reader is referred to \cite{DiepGiacomini} or to the original papers for a full description of the methods and results. Hereafter, we underline only the conditions for finding exact solutions, and enumerate the most striking features we found.  

Let us note that  different 2D Ising models
without crossing interactions can be mapped onto the
16-vertex model or the 32-vertex model, with the
free-fermion condition automatically satisfied in such
cases.  The partition function can be exactly found, leading to the critical condition allowing
to determine the phase transition as a function of interaction parameters.

\begin{figure}[th]              
\includegraphics[width=2.7in]{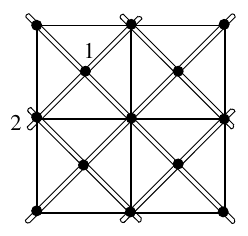}
\vspace*{8pt} \caption{ Centered square lattice. Interactions
between NN and NNN, $J_{1}$ and $J_{2}$, are denoted by double and
single bonds, respectively. The two sublattices are numbered 1 and
2. \label{re-fig17}}
\end{figure}

The  frustrated 2D Ising centered square lattice has been exactly solved by Vaks et al. \cite{Vaks} even before the notion of frustration was introduced \cite{Toulouse}   (see Fig.
\ref{re-fig17}  with NN and
NNN interactions, $J_{1}$ and $J_{2}$, respectively). The ground state
properties of this model are as follows : for $a = J_{2}/\mid
J_{1}\mid > - 1$, spins of sublattice 2 orders ferromagnetically
and the spins of sublattice 1 are parallel (antiparallel) to the
spins of sublattice 2 if $J_{1} > 0$ ( $< 0$ ); for $a < -1$,
spins of  sublattice 2 orders antiferromagnetically, leaving the
centered spins free to flip.  
%Fig5

\begin{figure}[th]             
\includegraphics[width=3.1 in]{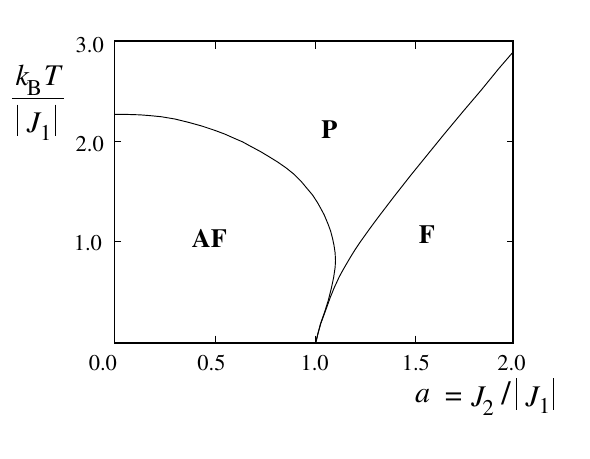}
\vspace*{4pt} \caption{ Phase diagram of centered square lattice \cite{Vaks}. See text for comments.}
 \label{re-fig18}
\end{figure}

The phase diagram of this model in the space $(a=J_2/|J_1|,T)$ is shown in Fig. \ref{re-fig18}. 
%Fig6

It is interesting to note that (i) in a small interval above $a=1$, when one increases $T$, the system
goes from the ferromagnetic (F) phase to the narrow paramagnetic phase between F and AF phases, then enters the partially ordered phase AF before going to the high-$T$ paramagnetic phase. In the AF phase the centered spins are free to flip as said above. The narrow paramagnetic phase between F and AF phases is called "reentrance phase" (a disordered phase between two ordered phases).

\begin{figure}[th]              %Fig~5
\includegraphics[width=2.2in]{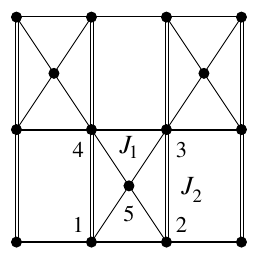}
\vspace*{8pt} \caption{ Kagom\'{e} lattice. Interactions between
NN and between NNN, $J_{1}$ and
$J_{2}$, are shown by single and double bonds, respectively.  The
lattice sites in a cell are numbered for decimation demonstration (not shown).
\label{re-fig7}}
\end{figure}

Let us consider as another example of the Ising model defined on a Kagom\'{e}
lattice, with two-spin interactions NN
and NNN, $J_{1}$ and $J_{2}$,
respectively, as shown in Fig. \ref{re-fig7}.

%Fig7

\begin{figure}[h]              
\includegraphics[width=3. in]{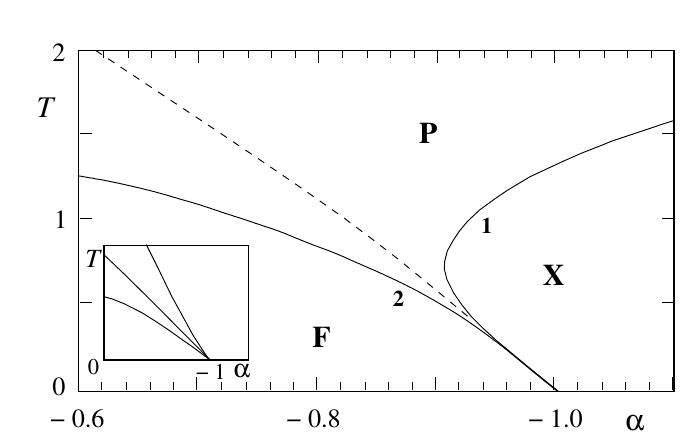}
\vspace*{8pt} \caption{ Phase diagram of the Kagom\'{e} lattice
with NN and NNN interactions in the region $J_{1} > 0$ of the space
($\alpha=J_{2}/J_{1}, T$). $T$ is measured in the unit of
$J_{1}/k_{B}$.  Solid lines are critical lines, dashed line is the
disorder line. P, F and X stand for paramagnetic, ferromagnetic
and partially disordered phases, respectively.  The inset shows
schematically enlarged region of the endpoint. \label{re-fig16}}
\end{figure}

The Hamiltonian is written as
\begin{equation}
H=-J_{1}\sum_{(ij)} \sigma_{i}\sigma_{j}-J_{2}\sum_{(ij)}
\sigma_{i}\sigma_{j}\label{re-eq19}
\end{equation}
%eq19
where
and the
first and second sums run over the spin pairs
connected by single and double bonds, respectively. Note that the model without  $J_2$ has been exactly solved a long time ago \cite{Ka/Na} showing no phase transition at
finite $T$ when $J_{1}$ is antiferromagnetic. Taking into account
the NNN interaction $J_{2}$ [see Fig.\ref{re-fig7} and Eq.
(\ref{re-eq19})], we have solved\cite{Aza87} this model by
transforming it into a 16-vertex model which satisfies the
free-fermion condition.  Let us just show the phase diagram in  the space ($J_2/J_1,T$) in Fig. \ref{re-fig16}.
There are three remarkable points: (i) the X phase is a partially disordered phase (only centered spins are disordered), (ii) the reentrance phase (paramagnetic phase between X and F phases near the endpoint, (iii) the existence of a disorder line starting from  the endpoint, separating the paramagnetic phase of F from that of  X (see signification  of the disorder line in \cite{Ste1,Ste2,Ste3}).    These points are very interesting,  they exist in highly frustrated solvable models but we believe that some real systems may bear some similar aspects. In the other models that we have studied \cite{Diep91b,Diep91a,Diep92}, these properties are  even more striking. The reader is referred to \cite{DiepGiacomini} for more discussion. 

It is interesting to note that the reentrance phase and the partial disorder found in the exactly solved models shown above exist also in quantum frustrated systems in three dimensions \cite{Quartu, Santamaria,Boubcheur} using Green's function and Monte Carlo methods.

%Fig8

\section{Fully frustrated systems and some other frustrated systems}\label{FFSystems}
\begin{figure}[ht]
\centering
\includegraphics[width=3.2 in]{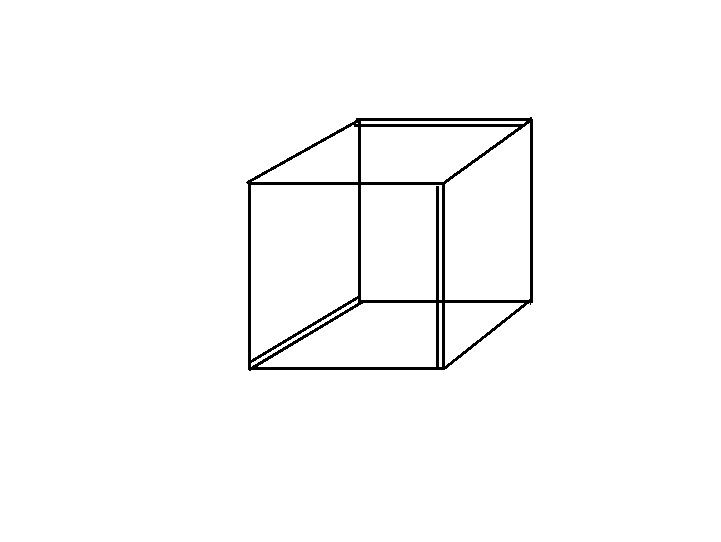}
%\begin{figure}[htb!]  %Fig4
%\centerline{\epsfig{file=IntroAFTL.eps,width=2.5in}}
\vspace{-1.5cm}
\caption{\label{FFSC} Unit cell of a fully SC  lattice. Double (simple) line is antiferromagnetic (ferromagnetic) interaction. Note that every  face of the cell is frustrated. The full lattice is made by stacking this cell in all  space directions. It is called fully frustrated SC lattice. }
\end{figure}

\begin{figure}
\centering
\includegraphics[width=3.2 in]{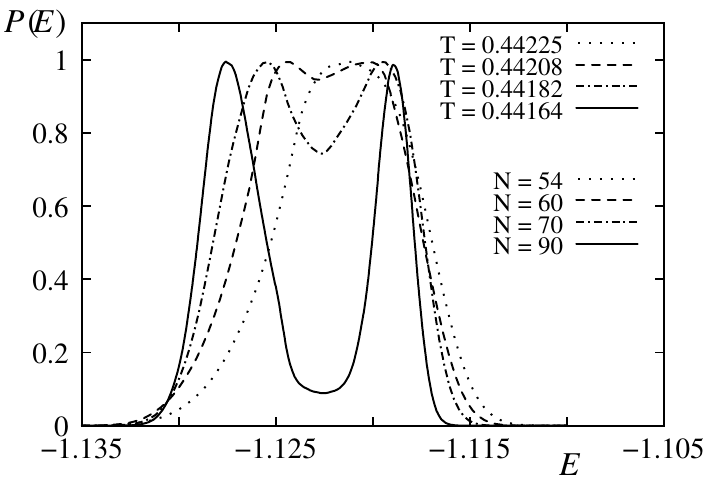}
\caption{Fully frustrated SC lattice with Heisenberg spins: Energy
histogram for several sizes $N=54$, 60, 70, 90 at $T_c$ indicated on the
figure. See text for comments. }\label{fig:PE}
\end{figure}

Several works in collaboration with G. Toulouse have been carried out on the so-called "fully frustrated simple cubic (SC) lattice" which is made by stacking the cubic cell shown in Fig. \ref{FFSC}  \cite{FFSC1,FFSC2,FFSC3,Derrida1,Derrida2}.  Let us say a few words on the work in which Toulouse was involved.  Classical vector spins on the sites of this  lattice interacting
via a periodic fully frustrated array of exchange interactions exhibit an interesting manifold of
GS configurations. In particular, for Heisenberg spins, the manifold has dimension 5, with
two continuous degeneracy parameters, in addition to global rotation angles. A configuration space
analysis, including pair and triangle overlap statistics, has been performed for this test model. The reader is referred to the original paper \cite{FFSC1} for details of the calculations.  We also studied the phase transition of this lattice with Ising \cite{FFSC2}, XY and Heisenberg spins \cite{FFSC3}. Note that in the XY case, using an algorithm which minimizes the local energies, we find twelve  periodic GS spin configurations.  A 12-fold degeneracy is very rare, if not unique, for a XY spin system. In the case of Heisenberg spins, the degeneracy is infinite. The phase transion in the case has been clarified after 30 years of uncertainty: using the high-performance Wang-Landau flat histogram MC algorithm \cite{WL1,WL2,brown,Schulz,Malakis} we have shown that the transition is of first order for both XY and Heisenberg spins \cite{NgoDiep2010,NgoDiep2011}: the energy histogram shown in Fig. \ref{fig:PE} has a double-peak structure at the transition temperature which is a signature of a first-order transition: the distance between the two peaks is the latent heat. 

%Fig9

%Fig10

Let us summarize  some frustrated systems having a first-order transition:

\begin{itemize}
\item Helimagnets with vector spins have a first-order transition \cite{Diep1992HCP,HoangDiep2012HCP}
\item Antiferromagnetic FCC lattice with Heisenberg spin has a first-order transition \cite{DiepKawamura}
\item $J_1-J_2$ SC model with Heisenberg spins \cite{PinettesDiep}, with Ising spins \cite{HoangDiep2011} both have first-order transition
\item The stacked triangular antiferromagnets: this model has been a fascinating subject since the 80's \cite{Kawamura1988,Kawamura1998}. Different methods give different results on the order of the phase transition. The very recent papers \cite{Reehorst,Delamotte2024} confirm the first-order  nature found by non-perturbative renormalization group \cite{Delamotte,Delamotte2010} and by Wang-Landau method \cite{NgoDiepHeis,NgoDiepXY}. The reader is referred to the papers \cite{Reehorst,Delamotte2024} for a complete bibliography on this subject.
\end{itemize}

\begin{figure}
\centering
\includegraphics[width=3.2 in]{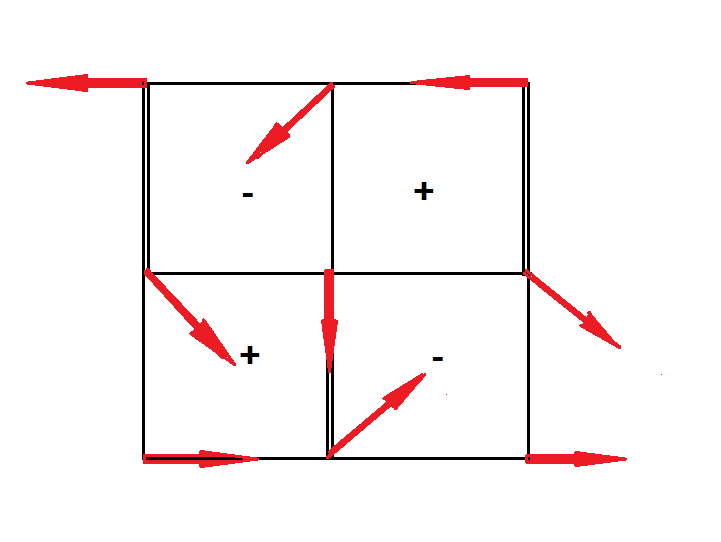}
\caption{The XY zigzag model: double and single lines are antiferromagnetic ($J=-1$) and ferromagnetic  ($J=1$), respectively. The angle between two neighboring spins linked by a ferromagnetic bond is 45$^\circ$ and that between spins linked by an antiferromagnetic bond is 135$^\circ$ [see the paragraph below Eq. (\ref{frust2a})]. The chiralities "+" and "-" are noted on the plaquettes of the spin configuration. See text for comments. }\label{zigzag}
\end{figure}

To conclude this section, let us  emphasize that the so-called zigzag model in  2D with XY spin model (Fig. \ref{zigzag}) shows a remarkable property: as the symmetry of the GS is two-fold (positive and negative chiralities), the breaking of this symmetry is of Ising type. However, the XY nature of the spins affects the criticality: we have shown in Ref. \cite{BoubcheurDiep1998} that (i)
the Ising and XY symmetries are broken at the same temperature (a single transition) (ii) the values of the
critical exponents are $\nu=0.852(2)$ and $\gamma=1.531(3)$, better
than most earlier MC results, in agreement with the suggestion
of a coupled ‘‘XY Ising" universality class by Lee et al. \cite{Nightingale,Granato,Lee,Granato1}. This is a new kind of criticality.

%Fig11

\section{Skyrmions}\label{Skyr}
Skyrmions are topologically stable spin structures observed in various materials and theories. They are objects of intensive studies since 2003 after the work of A. Bogdanov \cite{Bogdanov}.  We just cite a few works hereafter. The rapid development of the field of skyrmions is due to the potential applications using skyrmions in spintronics \cite{Fert2013,Leonov,Ezawa,Zhang,Zhang2017,Zhang2020,Zhang2021}.  

Skyrmions are generated by various interaction mechanisms. They can result from over-frustrated spin systems under an applied magnetic field \cite{Okubo,Hayami} or from the Dzyaloshinskii-Moriya interaction (DMI) as seen in \cite{ElHog2018,Sharafullin2019,Sharafullin2020,ElHog2022} among many other works in the literature. 
Note that a single skyrmion consists of a spin pointing down at the center. The surrounding spins turn around the center with incrreasing $z$ component when going away the center. At the boundary, all spins are up. 
There are two kinds of such arrangement. The first kind is the Bloch-type skyrmion in which the spins rotate in the tangential planes, namely perpendicular to the radial directions, when moving from the core to the periphery.  The second kind is the Neel-type skyrmion in which the spins rotate in the radial planes from the core to the periphery \cite{WangKang}.  In our model shown below, only the Bloch-type skyrmions are observed.

Hereafter, I take a recent work to illustrate some basic properties of a system with a DMI \cite{ElHog2022}. Consider  an antiferromagnetic triangular lattice with the following Hamiltonian

\begin{eqnarray}
\mathcal{H}&=&-J \sum_{\langle ij \rangle} \mathbf {S_i} \cdot \mathbf{S_j} -
 D \sum_{\langle ij \rangle} \mathbf u_{i,j} \cdot \mathbf {S_i} \times \mathbf {S}_{j} \nonumber\\
&&-H \sum_i S_i^z
\end{eqnarray}
where $\mathbf {S_i}$ is a classical Heisenberg spin of magnitude 1 occupying the lattice site $i$. The first sum runs over all spin nearest-neighbor (NN) pairs  with an antiferromagnetic exchange interaction $J$ ($J<0$), while the second sum is performed over all  DM interactions between NN.  $H$ is the magnitude of a magnetic field applied along the $z$ direction perpendicular to the lattice $xy$ plane.   The $D$-vector of the DMI is taken perpendicular to the $xy$ plane and is given by

\begin{eqnarray}
\mathbf D_{i,j}&=&D\mathbf u_{i,j}\label{D3a}\\
\mathbf D_{j,i}&=&D\mathbf u_{j,i}=-D\mathbf u_{i,j}\label{D3b}
\end{eqnarray}
where $\mathbf u_{i,j}$ is the unit vector on the $z$ axis, and $D$ represents the DM interaction strength.

\begin{figure}[h!]
\centering
\includegraphics[width=4cm]{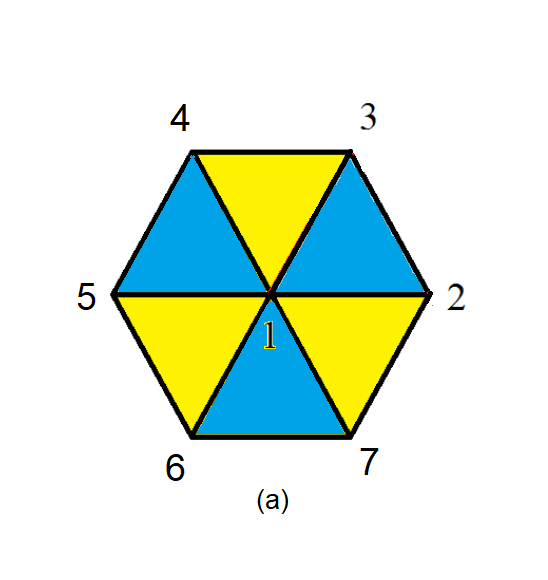}
\includegraphics[width=4cm]{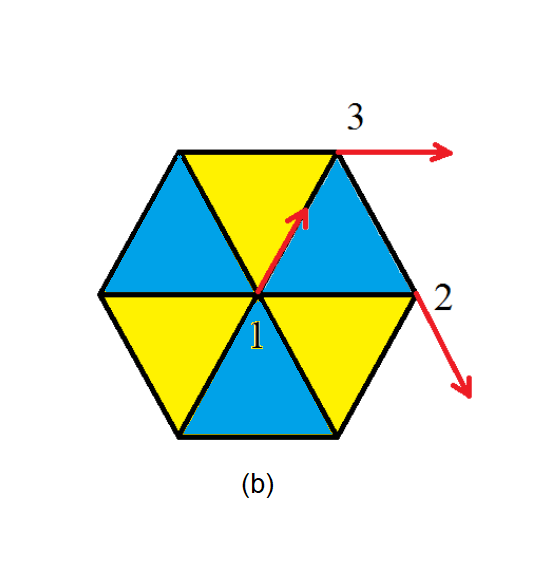}
\includegraphics[width=4.5cm]{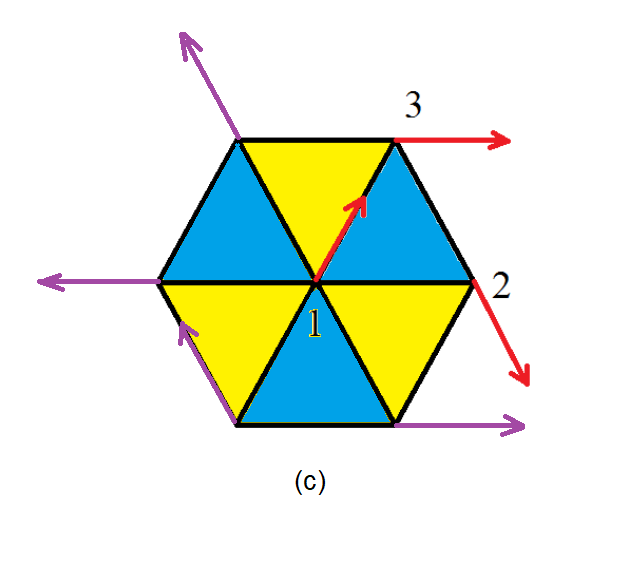}
\caption{Perpendicular $\mathbf D_{i,j}$: (a) The spins on a hexagon are numbered as indicated; (b)  Ground-state spin configuration with only Dzyaloshinskii-Moriya interaction on the triangular lattice ($J=0)$ is analytically determined. One angle is 120 degrees, the other two are 60 degrees. Note that the choice of the 120-degree angle  in this figure is along the horizontal spin pair. This configuration is one ground state, the other two ground states have the 120-degree angles on respectively the two diagonal spin pairs. Note also that the spin configuration is
invariant under the global spin rotation in the $xy$ plane; (c) example of construction of the GS for the whole hexagon: spins in violet are added in respecting the turning angle in each direction. See calcullations in Ref.    \cite{ElHog2022}.}\label{DMGSfig}
\end{figure}

The calculations of the GS have been done in details  in \cite{ElHog2022} for the case of perpendicular $\bf D$ where ($D\neq 0$,$J=0$, $H=0$), ($D\neq 0$,$J\neq 0$,$H=0$), and  ($D\neq 0$,$J\neq 0$,$H\neq 0$).  The first  two cases have been done by analytically minimizing the Hamiltonian and the last case by the numerical steepest descent method. 
The results of the first two cases are shown in Fig. \ref{DMGSfig}. 

%Fig12
%FigDMGS

\begin{figure}[h!]
%\vspace{-2cm}
\centering
\includegraphics[width=12cm]{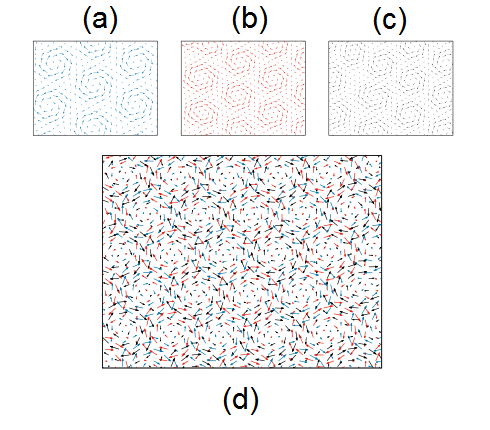}
%\vspace{-4cm}
\caption{In-plane $\mathbf D_{i,j}$:  The skyrmion crystal observed at $J$=-1, $D$=0.5 and $H$=3: A portion of skyrmion sublattice 1, 2 and 3 is respectively shown in (a), (b) and (c). Figure (d) shows the three interpenetrating antiferromagnetic skyrmion sublattices distinguished by three colors. }\label{ffig9}
\end{figure}

Now, we take the case of in-plane $\mathbf D_{i,j}$: we define $\mathbf D_{i,j}$ as

\begin{equation}\label{PDM}
\mathbf D_{i,j}=D(\mathbf r_j-\mathbf r_i)/|\mathbf r_j-\mathbf r_i|=D\mathbf {r}_{ij}
\end{equation}
where $D$ is a constant and $\mathbf {r}_{ij}$ denotes the unit vector along $\mathbf r_j-\mathbf r_i$.
The GS in this case is determined by the numerical steepest descent method. 
With ($D\neq 0$,$J\neq 0$) under an applied field $H$, the result is shown in Fig. \ref{ffig9} for $J=-1$, $D=0.5$ and $H=3$.  One observes the three interpenetrating skyrmion sublattices.   

%Fig13

These results are in agreement with early works using laborious MC simulations at low $T$ \cite{Rosales,Mohylna,Mohylna1,Liu}.

\begin{figure}[h]
\centering
%\vspace{2cm}
\includegraphics[width=12cm]{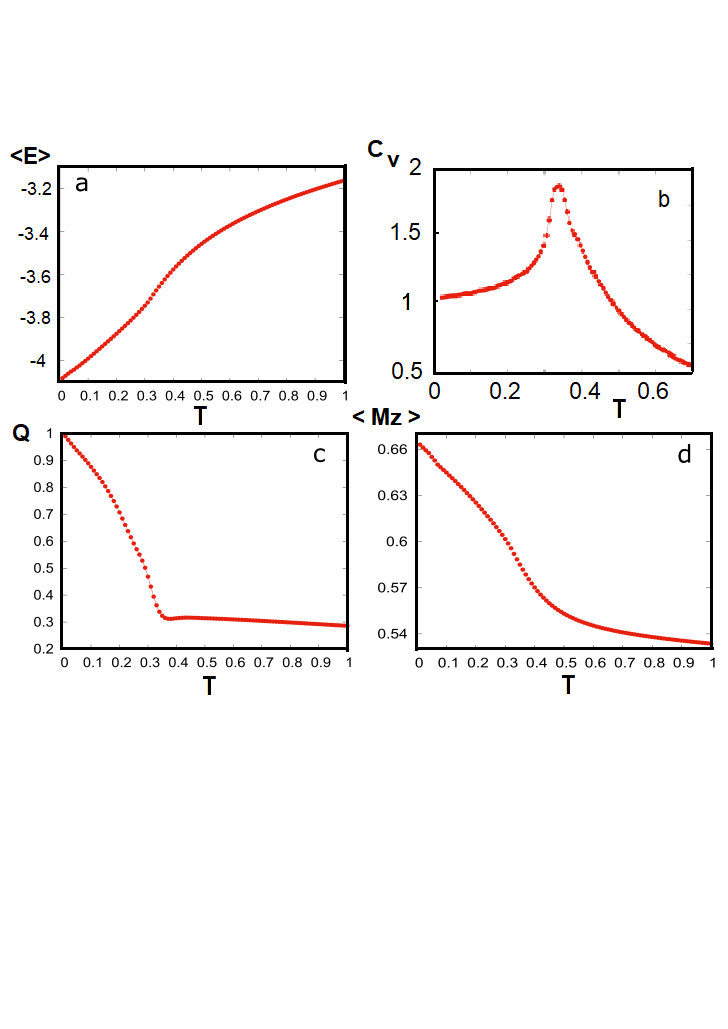}
\vspace{-5cm}
\caption{Results for $J$=-1, $D=0.5$, $H=3$: (a) Energy vs $T$, (b) Specific heat vs $T$, (c) Order parameter vs $T$, (d) $z$-component of the magnetization versus $T$. }\label{transit}
\end{figure}

To see if the skyrmions are stable at finite $T$, we have performed MC simulations.  In addition to the internal energy $E$ and the specific heat $C_v$, we calculated the order parameter $Q$ defined as the projection of the spin confuguration at the time $t$ at a given temperature $T$ on the GS. These quantities are given by

\begin{eqnarray}
\langle E \rangle &=& \langle \mathcal{H} \rangle /(2N) \\
C_v&=&\frac{\langle E^2 \rangle -\langle E \rangle^2}{k_BT^2}\\
Q(T)&=&\frac{1}{N(t_a-t_0)}\sum_i |\sum_{t=t_0}^{t_a} \mathbf S_i (T,t)\cdot \mathbf S_i(T=0)|\label{OrderP}
\end{eqnarray}
where $t_a-t_0$ is the averaging time.   Note that the order parameter $Q$ mimics the Edwards-Anderson order parameter \cite{EAmodel} defined for random spin configurations (spin glass). The results are shown in Fig. \ref{transit} where the $z$ component $M_z$ is also displayed.  We observe a transition from the skyrmion phase to the paramagnetic phase at $T_c/|J|\simeq 0.35$ where $C_v$ shows a peak, and where $Q$ vanishes (the non zero values of $Q$ and $M_z$ after $T_c$ are due to the applied field).  The stability of skyrmions at finite temperatures is essential because devices based on skyrmions  are helpful only if they work at finite $T$.   

Note that the size of the skyrmion does not depend on the lattice size when it is larger than 40x40 (smaller sizes induce some boundary effects). The skyrmion size depends however on the strength of the applied field $H$: the stronger $H$, the smaller the diameter of the skyrmion. The results shown in Fig. \ref{transit} were obtained with the size 100x100. The size effects for larger sizes are insignificant. As the last remark, we do not know at the time being about the nature of the transition observed here. A question which naturally arises is that whether or not it follows the standard finite-size scaling theory of a second-order phase transition (see \cite{Zinn,Cardy} for a summary of this theory). It needs a further study to clarifdy this point.

%Fig14

To conclude this section, we note that what shown above is some basic properties of skyrmions. The dynamics of skyrmions which is the heart of applications in spintronics, is not presented here to keep a reasonable length of the paper. We note however that the small size of skyrmions (a dozen of spins), their topologically protected stability and their easy mobility \cite{Zhang,Zhang2020,WangKang} make them very suitable for applications.
The reader is referred to more  references cited in this paper on this subject, from Ref. \cite{Bogdanov} to Ref. \cite{Liu} (the titles tell much about their contents).

\section{Quantum Theory for Non-Collinear Spin Configurations: Green's Function Method}\label{Green}

The study of elementary excitations in spin systems is one of the main subjects in  magnetism \cite {DiepTM}. In magnetic materials, elementary excitations, called "spin waves" or "magnons", dominate the low-temperature properties.  Various means of  investigation, theoretically and experimentally, such as spin-wave theories and inelastic neutron scattering, have been developped since the 60's for magnetic materials. Theoretically, the theory of magnons based on the Holstein-Primakoff was the most advanced theory, but the validity of this method is limited to the region of very low temperatures \cite{DiepTM}. The first Green's function method introduced by Zubarev in 1960 \cite{Zubarev} allowed us to calculate bulk magnetic properties up to the transition temperature $T_C$.  I have applied this method with success in a study of magnetic thin films in 1979 \cite{Diep1979}.  Note that these studies have been limited to the cases where the magnetic ordering is collinear.
I have developped with R. Quartu a Green's function technique in 1997 to study non-colinear spin configurations \cite{Quartu} in frustrated spin systems: we  observe that the commmutation relations between spin-deviation operators are valid only when the spin lies on its quantization axis.  However, in a non-collinear configuration, each spin has its own quantization axis. Our idea was, for a given spin, to use its "local coordinates".  The calculation was possible for any non-collinear  spin configuration, as seen in our publications afterward: one can mention the case of frustrated surface in thin films \cite{Ngo2007a,Ngo2007b} or helimagnetic thin films  \cite{DiepHeli2015} where the surface spin reconstruction is strong. Surface effects in thin films have been widely studied theoretically, experimentally, and
numerically, during the last three decades \cite{Bland,Zangwill}. Nevertheless,
surface effects in frustrated systems such as helimagnets have only been recently
studied: surface spin structures \cite{Mello}, MC
simulations \cite{Cinti}, magnetic field effects on the phase diagram
in Ho \cite{Rodrigues}, and a few experiments \cite{Karhu1,Karhu2}. Helical magnets present potential applications in spintronics
with predictions of spin-dependent electron transport in
these magnetic materials \cite{Heurich,Wessely,Jonietz}.  In the following, I recall the main steps of the calculation, the reader is referred to the original paper \cite{DiepHeli2015} for more details.

\subsection{Green's Function Method for Helimagnets}
To illustrate the method, let us consider the simplest case of helimagnets: the turn angle takes place in the $c$-direction of a body-centered tetragonal (BCT) lattice, perpendicular to the film. This is due to the NNN antiferromagnetic interaction $J_2<0$ in that direction. For NN interaction, we assume a ferromagnetic interaction $J_1>0$. The film thickness consists of $N_z$ layer. In the $xy$ plane, we use the periodic boundary conditions. The hamiltonian is given by

\begin{equation}
\mathcal H_e=-\sum_{\left<i,j\right>}J_{1}\mathbf S_i\cdot\mathbf
S_j  -\sum_{\left<i,j\right>}J_{2}\mathbf S_i\cdot\mathbf
S_j  \label{eqn:hamil1}
\end{equation}
where $\mathbf S_i$ and $\mathbf S_j$ are two quantum Heisenberg spins occupying the lattice sites $i$ and $j$. The first and second sums run over the NN  and NNN spin pairs, respectively. 

For the bulk BCT lattice, modifying the method to obtain Eq. (\ref{helichain}) for a chain, we obtain the critical value $|J_2|/J_1>1$ above which the GS is helical. 

Now, in a film there is a lack of NN and NNN at the two surfaces. This modifies the turn angle at and near the surfaces. The reader is referreed to Ref. \cite{DiepHeli2015} for the calculations.  The results are shown in Table
\ref{table}

%\begin{widetext}
\begin{center}
\begin{table}
\begin{tabular}{|l|c|c|c|c|r|}
\hline
$J_2/J_1$ & $\cos \theta_{1,2}$ & $\cos\theta_{2,3}$ & $\cos\theta_{3,4}$ & $\cos\theta_{4,5}$ & $\alpha$(bulk)  \\
\hline
&&&&&\\
-1.2  &  0.985($9.79^\circ$) &  0.908($24.73^\circ$)       &    0.855($31.15^\circ$)    &   0.843($32.54^\circ$)  &   $33.56^\circ$    \\
-1.4 &  0.955($17.07^\circ$)&  0.767($39.92^\circ$)  &  0.716($44.28^\circ$)    &   0.714($44.41^\circ$)   &  $44.42^\circ$     \\
-1.6 & 0.924($22.52^\circ$) &  0.633($50.73^\circ$) & 0.624($51.38^\circ$)  &  0.625($51.30^\circ$)   &  $51.32^\circ$        \\
-1.8     &  0.894($26.66^\circ$)  &  0.514($59.04^\circ$)  & 0.564($55.66^\circ$)   &  0.552($56.48^\circ$)  &  $56.25^\circ$    \\
-2.0       &  0.867($29.84^\circ$)  &  0.411($65.76^\circ$) &  0.525($58.31^\circ$)   &  0.487($60.85^\circ$)  & $60^\circ$   \\
&&&&&\\
\hline
\end{tabular}
\vspace{1cm}
\caption{Values of $\cos \theta_{n,n+1}=\alpha_n$ between two adjacent layers are shown for various
values of $J_2/J_1$. Only angles of the first half of the 8-layer film are shown: other angles are, by symmetry,
$\cos\theta_{7,8}$=$\cos\theta_{1,2}$, $\cos\theta_{6,7}$=$\cos\theta_{2,3}$, $\cos\theta_{5,6}$=$\cos\theta_{3,4}$. The values in parentheses are angles in degrees.
The last column shows the value of the angle in the bulk case (infinite thickness).
For presentation, angles are shown with two  digits. \label{table} }
\end{table}
\end{center}
%\end{widetext}

\vspace{1cm}
In the following, using the spin configuration obtained at each $J_2/J_1$ we calculate the spin-wave excitation and properties
of the film such as the zero-point quantum spin fluctuations, the layer magnetizations versus $T$ and the critical temperature $T_C$.

\begin{figure}[htb]
\centering
\includegraphics[width=6cm]{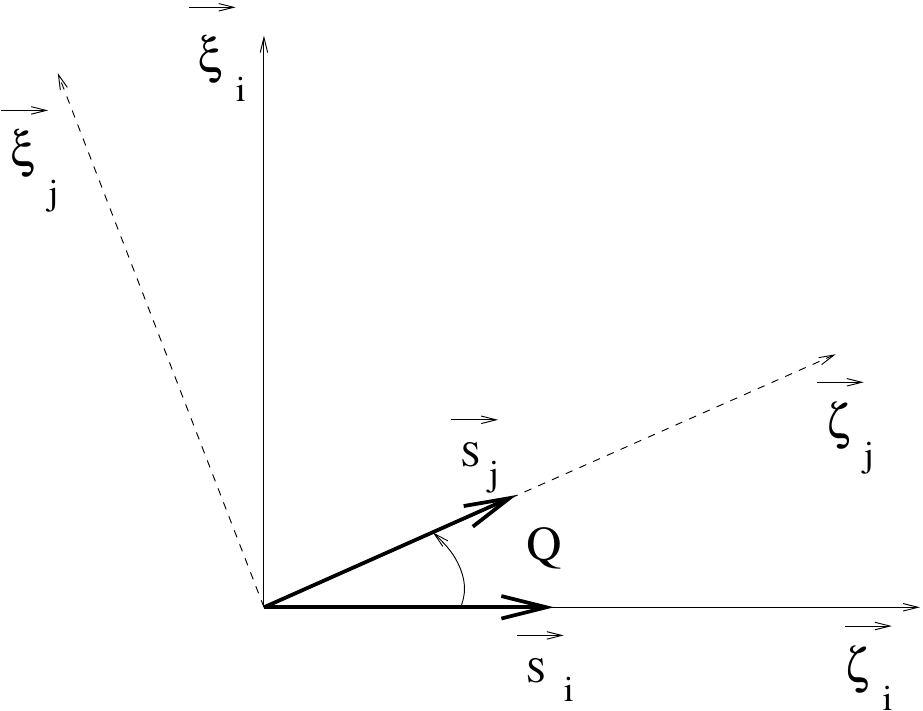}  % .eps
\caption{ Local coordinates in a $xy$-plane perpendicular to the $c$-axis. $Q$ denotes $\theta_{j}-\theta_i$.}\label{local}
\end{figure}

As said earlier, we have to use the local coordinates where a given spin should lie on its quantization axis. Needless to  say, in a non-collinear spin configfuration, each spin has its own quantization axis. Let us show for a spin pair ( $\mathbf S_i$, $\mathbf S_j$) the local coordinates shown in Fig. \ref{local}:  the quantization axis of spin $\vec S_i$ is on
its $\zeta_i$ axis which lies in the plane, the $\eta_i$ axis of $\vec S_i$ is along the $c$-axis, and the $\xi_i$
axis forms with $\eta_i$ and $\zeta_i$ axes a direct trihedron.

%Fig15

Since the spin configuration is planar, all spins
have the same $\eta$ axis. Furthermore,  all spins in a given layer are parallel.  Let $\hat \xi_i$, $\hat \eta_i$
and $\hat \zeta_i$ be the unit vectors on the local $(\xi_i,\eta_i,\zeta_i)$ axes. We write
\begin{eqnarray}
\vec S_i&=&S_i^x\hat \xi_i+S_i^y\hat \eta_i+S_i^z\hat \zeta_i\label{SI}\\
\vec S_j&=&S_j^x\hat \xi_j+S_j^y\hat \eta_j+S_j^z\hat \zeta_j\label{SJ}
\end{eqnarray}
We have (see Fig. \ref{local})
\begin{eqnarray}
\hat \xi_j&=&\cos \theta_{ij}\hat \zeta_i+\sin \theta_{ij}\hat \xi_i\nonumber\\
\hat \zeta_j&=&-\sin \theta_{ij}\hat \zeta_i+\cos \theta_{ij}\hat \xi_i\nonumber\\
\hat \eta_j&=&\hat \eta_i\nonumber
\end{eqnarray}
where $\cos \theta_{ij}=\cos (\theta_i-\theta_j)$ is the angle between two
spins $i$ and $j$.
Replacing these into Eq. (\ref{SJ}) to express $\vec S_j$ in the $(\hat \xi_i,\hat \eta_i,\hat \zeta_i)$ coordinates, then calculating
$\vec S_i \cdot \vec S_j$, we obtain the following exchange Hamiltonian from (\ref{eqn:hamil1}):
\begin{eqnarray}
\mathcal H_e &=& - \sum_{<i,j>}
J_{i,j}\Bigg\{\frac{1}{4}\left(\cos\theta_{ij} -1\right)
\left(S^+_iS^+_j +S^-_iS^-_j\right)\nonumber\\
&+& \frac{1}{4}\left(\cos\theta_{ij} +1\right) \left(S^+_iS^-_j
+S^-_iS^+_j\right)\nonumber\\
&+&\frac{1}{2}\sin\theta_{ij}\left(S^+_i +S^-_i\right)S^z_j
-\frac{1}{2}\sin\theta_{ij}S^z_i\left(S^+_j
+S^-_j\right)\nonumber\\
&+&\cos\theta_{ij}S^z_iS^z_j\Bigg\}
\label{eq:HGH2}
\end{eqnarray}
where the sum is performed on both NN and NNN spin pairs with corresponding exchange interactions $J_{ij}$ and corresponding angles $\theta_{ij}$.

Let us note that according to the theorem of Mermin and Wagner \cite{Mermin}
 continuous isotropic spin models such as XY and Heisenberg spins
do not have long-range ordering at finite temperatures in two dimensions. The Heisenberg model in a thin film,  of small thickness, though not in 2D, may need a small anisotropic interaction  to enhance its long-range ordering at finite temperatures. We add therefore the following anisotropic interaction along the in-plane local spin-quantization axes $z$ of $\mathbf S_i$
and $\mathbf S_j$:
\begin{equation}
\mathcal H_a= -\sum_{<i,j>} I_{i,j}S^z_iS^z_j\cos\theta_{ij}
\end{equation}
where $I_{i,j}(>0)$ is supposed to be positive, very small compared to $J_1$, and limited to NN on the $c$-axis.
The full Hamiltonian is thus
$\mathcal H=\mathcal H_e+\mathcal H_a$.

We define the following two double-time Green's functions in the real space:
\begin{eqnarray}
G_{i,j}(t,t')&=&<<S_i^+(t);S_{j}^-(t')>>\nonumber\\
&=&-i\theta (t-t')
<\left[S_i^+(t),S_{j}^-(t')\right]> \label{green59a}\\
F_{i,j}(t,t')&=&<<S_i^-(t);S_{j}^-(t')>>\nonumber\\
&=&-i\theta (t-t')
<\left[S_i^-(t),S_{j}^-(t')\right]>\label{green60}
\end{eqnarray}
We need these two functions because the equation of motion of the first function generates functions of
the second type, and vice-versa. These equations of motion are
\begin{eqnarray}
i\hbar\frac {d}{dt}G_{i,j}\left( t,t'\right) &=& \left<\left[ S^+_i
\left( t\right) , S^-_j \left( t'\right)\right]\right>\delta\left(
t-t'\right) \nonumber\\
&-& \left<\left< \left[\mathcal H, S^+_i\left( t\right)\right] ;
S^-_j \left( t'\right) \right>\right>,
\label{eq:HGEoMG}\\
i\hbar\frac {d}{dt}F_{i,j}\left( t,t'\right) &=& \left<\left[ S^-_i
\left( t\right) , S^-_j \left( t'\right)\right]\right>\delta\left(
t-t'\right)\nonumber \\
&-& \left<\left< \left[\mathcal H, S^-_i\left( t\right)\right] ;
S^-_j \left( t'\right) \right>\right>, \label{eq:HGEoMF}
\end{eqnarray}
Expanding the commutators, and using the Tyablikov decoupling scheme \cite{Tyablikov,Tyablikov1} for
higher-order functions, for example $<<S_{i'}^zS_i^+(t);S_{j}^-(t')>>\simeq <S_{i'}^z><<S_i^+(t);S_{j}^-(t')>>$ etc., we
obtain the following general equations for non collinear magnets:
 \begin{eqnarray}
 i\hbar \frac{dG_{i,j}(t,t')}{dt}&=&2<S_i^z>\delta_{i,j} \delta (t-t')\nonumber\\
 &-&\sum_{i'}J_{i,i'}[<S_i^z>(\cos \theta_{i,i'}-1)\times \nonumber\\
 &\times& F_{i',j}(t,t')\nonumber\\
&+&<S_i^z>(\cos \theta_{i,i'}+1)G_{i',j}(t,t')\nonumber\\
&-&2<S_{i'}^z>\cos \theta_{i,i'}G_{i,j}(t,t')]\nonumber\\
&+&2\sum_{i'}I_{i,i'}<S_{i'}^z>\cos \theta_{i,i'}G_{i,j}(t,t')\nonumber\\
&&\label{GFG0}\\
i\hbar \frac{dF_{i,j}(t,t')}{dt}&=&\sum_{i'}J_{i,i'}[<S_i^z>(\cos \theta_{i,i'}-1)\times\nonumber\\
&\times& G_{i',j}(t,t')\nonumber\\
&+&<S_i^z>(\cos \theta_{i,i'}+1)F_{i',j}(t,t')\nonumber\\
&-&2<S_{i'}^z>\cos \theta_{i,i'}F_{i,j}(t,t')]\nonumber\\
&-&2\sum_{i'}I_{i,i'}<S_{i'}^z>\cos \theta_{i,i'}F_{i,j}(t,t')\nonumber\\
&&\label{GFF0}
\end{eqnarray}

In the case of a BCC thin film with a (001) surface considered here,  the above equations yield a closed system of
coupled equations within the Tyablikov decoupling scheme.
For clarity, we  separate the sums on NN interactions and NNN interactions as follows:
 \begin{eqnarray}
 i\hbar \frac{dG_{i,j}(t,t')}{dt}&=&2<S_i^z>\delta_{i,j} \delta (t-t')\nonumber\\
 &-&\sum_{k'\in NN}J_{i,k'}[<S_i^z>(\cos \theta_{i,k'}-1) \times \nonumber\\
 &\times& F_{k',j}(t,t')\nonumber\\
 &+&<S_i^z>(\cos \theta_{i,k'}+1) G_{k',j}(t,t')\nonumber\\
 &-&2<S_{k'}^z>\cos \theta_{i,k'}G_{i,j}(t,t')]\nonumber\\
 &+&2\sum_{k'\in NN}I_{i,k'}<S_{k'}^z>\cos \theta_{i,k'}G_{i,j}(t,t')\nonumber\\
&-&\sum_{i'\in NNN}J_{i,i'}[<S_i^z>(\cos \theta_{i,i'}-1) \times \nonumber\\
&\times& F_{i',j}(t,t')\nonumber\\
&+&<S_i^z>(\cos \theta_{i,i'}+1) G_{i',j}(t,t')\nonumber\\
&-&2<S_{i'}^z>\cos \theta_{i,i'}G_{i,j}(t,t')]\label{GFG2}\\
i\hbar \frac{dF_{k,j}(t,t')}{dt}&=&\sum_{i'\in NN}J_{k,i'}[<S_k^z>(\cos \theta_{k,i'}-1)\times\nonumber\\
&\times&  G_{i',j}(t,t')\nonumber\\
&+&<S_k^z>(\cos \theta_{k,i'}+1) F_{i',j}(t,t')\nonumber\\
&-&2<S_{i'}^z>\cos \theta_{k,i'}F_{k,j}(t,t')]\nonumber\\
&-&2\sum_{i'\in NN}I_{k,i'}<S_{i'}^z>\cos \theta_{k,i'}F_{k,j}(t,t')\nonumber\\
&+&\sum_{k'\in NNN}J_{k,k'}[<S_k^z>(\cos \theta_{k,k'}-1) \times \nonumber\\
&\times& G_{k',j}(t,t')\nonumber\\
&+&<S_k^z>(\cos \theta_{k,k'}+1) F_{k',j}(t,t')\nonumber\\
&-&2<S_{k'}^z>\cos \theta_{k,k'}F_{k,j}(t,t')]\label{GFF2}
\end{eqnarray}
For simplicity, except otherwise stated, all NN interactions $(J_{k,k'}, I_{k,k'})$ are taken equal to $(J_1,I_1)$ and all NNN interactions
are taken equal to $J_2$ in the following.  Furthermore, let us define the film coordinates which are used
below: the $c$-axis is called $z$-axis, planes parallel to the film surface are called $xy$-planes and the Cartesian components
of the spin position $\mathbf R_i$ are denoted by $(\ell_i,m_i,n_i)$.

We now introduce the following in-plane Fourier
transforms:

\begin{eqnarray}
G_{i, j}\left( t, t'\right) &=& \frac {1}{\Delta}\int\int_{BZ} d\mathbf
k_{xy}\frac{1}{2\pi}\int^{+\infty}_{-\infty}d\omega e^{-i\omega
\left(t-t'\right)}\nonumber\\
&&\hspace{0.7cm}\times g_{n_i,n_j}\left(\omega , \mathbf k_{xy}\right)
e^{i\mathbf k_{xy}\cdot \left(\mathbf R_i-\mathbf
R_j\right)},\label{eq:HGFourG}\\
F_{k, j}\left( t, t'\right) &=& \frac {1}{\Delta}\int\int_{BZ} d\mathbf
k_{xy}\frac{1}{2\pi}\int^{+\infty}_{-\infty}d\omega e^{-i\omega
\left(t-t'\right)}\nonumber\\
&&\hspace{0.7cm}\times f_{n_k,n_j}\left(\omega , \mathbf k_{xy}\right)
e^{i\mathbf k_{xy}\cdot \left(\mathbf R_k-\mathbf
R_j\right)},\label{eq:HGFourF}
\end{eqnarray}
where $\omega$ is the spin-wave frequency, $\mathbf k_{xy}$
denotes the wave-vector parallel to $xy$ planes and $\mathbf R_i$ is
the position of the spin at the site $i$. $n_i$, $n_j$ and $n_k$ are
respectively the $z$-component indices of the layers where the sites $\mathbf R_i$,  $\mathbf R_j$ and $\mathbf R_k$
belong to. The integral over $\mathbf k_{xy}$ is performed in the
first Brillouin zone ($BZ$) whose surface is $\Delta$ in the $xy$
reciprocal plane.  For convenience, we denote $n_i=1$ for all sites on the surface layer, $n_i=2$ for all sites of the second layer and so on.

Note that for a three-dimensional case,  making a 3D Fourier transformation  of Eqs. (\ref{GFG2})-(\ref{GFF2})
we obtain the spin-wave dispersion relation in the absence of anisotropy:
\begin{equation}
\hbar\omega=\pm \sqrt{A^2-B^2}
\end{equation}
where
\begin{eqnarray}
A&=& J_1 \left< S^z\right>[\cos\theta+1]Z\gamma+2Z J_1\left< S^z\right> \cos \theta\nonumber\\
&&+J_2 \left< S^z\right> [\cos (2\theta)+1]Z_c\cos (k_za) \nonumber\\
&&+2Z_cJ_2 \left< S^z\right> \cos (2\theta)\nonumber\\
B&=& J_1 \left< S^z\right> (\cos\theta -1)Z\gamma\nonumber\\
&&+J_2 \left< S^z\right>[\cos(2\theta) -1]Z_c\cos (k_za) \nonumber
\end{eqnarray}
where $Z=8$ (NN number), $Z_c=2$ (NNN number on the $c$-axis), $\gamma=\cos (k_xa/2)\cos (k_ya/2)\cos (k_za/2)$ ($a$: lattice constant).
We see that $\hbar\omega$ is zero when $A=\pm B$, namely at $k_x=k_y=k_z=0$ ($\gamma=1$) and at $k_z=2\theta$ along the helical axis.
The case of ferromagnets (antiferromagnets) with NN interaction only is recovered by putting $\cos \theta=1$ $(-1)$ \cite{Diep1979}.

Let us return to the film case. We make the in-plane Fourier transformation Eqs. (\ref{eq:HGFourG})-(\ref{eq:HGFourF})
for  Eqs. (\ref{GFG2})-(\ref{GFF2}). We obtain the following matrix equation
\begin{equation}
\mathbf M \left( \omega \right) \mathbf h = \mathbf u,
\label{eq:HGMatrix}
\end{equation}
where $\mathbf M\left(\omega\right)$ is a square matrix of dimension
$\left(2N_z \times 2N_z\right)$, $\mathbf h$ and $\mathbf u$ are
the column matrices which are defined as follows
\begin{equation}
\mathbf h = \left(%
\begin{array}{c}
  g_{1,n'} \\
  f_{1,n'} \\
  \vdots \\
  g_{n,n'} \\
  f_{n,n'} \\
    \vdots \\
  g_{N_z,n'} \\
  f_{N_z,n'} \\
\end{array}%
\right) , \mathbf u =\left(%
\begin{array}{c}
  2 \left< S^z_1\right>\delta_{1,n'}\\
  0 \\
  \vdots \\
  2 \left< S^z_{N_z}\right>\delta_{N_z,n'}\\
  0 \\
\end{array}%
\right) , \label{eq:HGMatrixgu}
\end{equation}
where,  taking $\hbar=1$ hereafter,
%\begin{widetext}
\begin{equation}
\mathbf M\left(\omega\right) = \left(%
\begin{array}{cccccccccccc}
  \omega+A_1&0    & B^+_1    & C^+_1& D_1^+& E_1^+& 0&0&0&0&0&0\\
   0    & \omega-A_1  & -C^+_1 & -B^+_1 &-E_1^+&-D_1^+&0&0&0&0&0&0\\
   \cdots & \cdots & \cdots &\cdots&\cdots&\cdots&\cdots&\cdots&\cdots&\cdots&\cdots&\cdots\\
 \cdots&D_n^-&E_n^-&B^-_{n}&C^-_{n}&\omega+A_{n}&0&B^+_{n}&C^+_{n}&D_n^+&E_n^+&\cdots\\
 \cdots&-E_n^-&-D_n^-&-C^-_{n}&-B^-_{n}&0&\omega-A_{n}&-C^+_{n}&-B^+_{n}&-E_n^+&-D_n^+&\cdots\\
         \cdots  & \cdots & \cdots & \cdots &\cdots&\cdots&\cdots&\cdots&\cdots&\cdots&\cdots&\cdots \\
  0& 0&0&0& 0& 0& D^-_{N_z}& E^-_{N_z}  & B^-_{N_z}   & C^-_{N_z}   &\omega + A_{N_z}&0\\
  0&0&0&0&0&0&-E^-_{N_z}& -D^-_{N_z} & -C^-_{N_z}  & -B^-_{N_z}&0  & \omega-A_{N_z}\\
\end{array}%
\right) \label{eq:HGMatrixM}
\end{equation}
%\end{widetext}
where
\begin{eqnarray}
A_{n} &=& - 8J_1(1+d) \Big[\left< S^z_{n+1}\right>
\cos\theta_{n,n+1}\nonumber\\
&&+\left< S^z_{n-1}\right>
\cos\theta_{n,n-1}\Big]\nonumber\\
&-&2J_2 \Big[\left< S^z_{n+2}\right>
\cos\theta_{n,n+2}\nonumber\\
&&+\left< S^z_{n-2}\right>
\cos\theta_{n,n-2}\Big]\nonumber
\end{eqnarray}
where $n=1,2,...,N_z$, $d=I_1/J_1$, and
\begin{eqnarray}
B_n^\pm &=& 4J_1 \left< S^z_{n}\right>(\cos\theta_{n,n\pm 1}+1)\gamma \nonumber\\
C_n^\pm &=& 4J_1 \left< S^z_{n}\right>(\cos\theta_{n,n\pm 1}-1)\gamma \nonumber\\
E_n^\pm &=& J_2 \left< S^z_{n}\right>(\cos\theta_{n,n\pm 2}-1)\nonumber\\
D_n^\pm &=& J_2 \left< S^z_{n}\right>(\cos\theta_{n,n\pm 2}+1) \nonumber
\end{eqnarray}
In the above expressions, $\theta_{n,n\pm
1}$ the angle between a spin in the layer $n$ and its NN spins in layers $n\pm 1$ etc. and
$\gamma = \cos \left( \frac{k_x a}{2} \right)\cos \left( \frac{k_y a}{2} \right).$

Solving det$|\mathbf M|=0$, we obtain the spin-wave spectrum
$\omega$ of the present system: for each value ($k_x,k_y)$, there are 2$N_z$ eigen-values of $\omega$ corresponding to two opposite spin precessions as in antiferromagnets (the dimension of det$|\mathbf M|$ is $2N_z\times 2N_z$).  Note that the above equation depends on the values of $<S_n^z>$ ($n=1,...,N_z$).
Even at temperature $T=0$, these $z$-components are not equal to $1/2$ because we are dealing with an antiferromagnetic
system where fluctuations at $T=0$ give rise to the so-called zero-point spin contraction \cite{DiepTM}. Worse, in our system
with the existence of the film surfaces, the spin contractions are not spatially uniform as will be seen below.
So the solution of det$|\mathbf M|=0$ should be found by iteration.  This will be explicitly shown hereafter.

The solution for $g_{n,n}$ is given by
\begin{equation}
g_{n,n}(\omega) = \frac{\left|\mathbf M\right|_{2n-1}}{\left|\mathbf M\right|},
\end{equation}
where $\left|\mathbf M\right|_{2n-1}$ is the determinant made by
replacing the $2n-1$-th column of $\left|\mathbf M\right|$ by
$\mathbf u$ given by Eq. (\ref{eq:HGMatrixgu}) [note that $g_{n,n}$ occupies the $(2n-1)$-th line of the matrix $\mathbf h$]. Writing now
\begin{equation}
\left|\mathbf M\right| = \prod_i \left[\omega -
\omega_i\left(\mathbf k_{xy}\right)\right],
\end{equation}
we see that $\omega_i\left(\mathbf k_{xy}\right) ,\ i = 1,\cdots
,\ 2N_z$, are poles of  $g_{n,n}$.
$\omega_i\left(\mathbf k_{xy}\right)$ can be obtained by solving
$\left|\mathbf M\right|=0$. In this case, $g_{n,n}$ can be
expressed as
\begin{equation}
g_{n, n}(\omega) = \sum_i\frac {D_{2n-1}\left(\omega_i\left(\mathbf
k_{xy}\right)\right)}{\left[ \omega - \omega_i\left(\mathbf
k_{xy}\right)\right]}, \label{eq:HGGnn}
\end{equation}
where $D_{2n-1}\left(\omega_i\left(\mathbf k_{xy}\right)\right)$ is
\begin{equation}
D_{2n-1}\left(\omega_i\left(\mathbf k_{xy}\right)\right) = \frac{\left|
\mathbf M\right|_{2n-1} \left(\omega_i\left(\mathbf
k_{xy}\right)\right)}{\prod_{j\neq i}\left[\omega_j\left(\mathbf
k_{xy}\right)-\omega_i\left(\mathbf k_{xy}\right)\right]}.
\end{equation}

Next, using the spectral theorem which relates the correlation
function \(\langle S^-_i S^+_j\rangle \) to the Green's function \cite{Zubarev}, we have
\begin{eqnarray}
\left< S^-_i S^+_j\right> &=& \lim_{\varepsilon\rightarrow 0}
\frac{1}{\Delta}\int\int d\mathbf k_{xy}
\int^{+\infty}_{-\infty}\frac{i}{2\pi}\big( g_{n, n'}\left(\omega
+ i\varepsilon\right)\nonumber\\
&-& g_{n, n'}\left(\omega - i\varepsilon\right)\big)
\frac{d\omega}{e^{\beta\omega} - 1}e^{i\mathbf
k_{xy}\cdot\left(\mathbf R_i -\mathbf R_j\right)},
\end{eqnarray}
where $\epsilon$ is an  infinitesimal positive constant and
$\beta=(k_BT)^{-1}$, $k_B$ being the Boltzmann constant.

Using the Green's function presented above, we can calculate
self-consistently various physical quantities as functions of
temperature $T$.  The magnetization $\langle S_{n}^z\rangle$ of the $n$-th layer is given by
\begin{eqnarray}
\langle S_{n}^z\rangle&=&\frac{1}{2}-\left< S^-_{n} S^+_{n}\right>\nonumber\\
&=&\frac{1}{2}-
   \lim_{\epsilon\to 0}\frac{1}{\Delta}
   \int
   \int d{\bf k_{xy}}
   \int\limits_{-\infty}^{+\infty}\frac{i}{2\pi}
   [ g_{n,n}(\omega+i\epsilon)\nonumber\\
   &&-g_{n,n}(\omega-i\epsilon)]
\frac{d\omega}{\mbox{e}^{\beta \omega}-1}\label{lm1}
\end{eqnarray}
Replacing Eq. (\ref{eq:HGGnn}) in Eq. (\ref{lm1}) and making use of the following identity

\begin{equation}\label{id}
\frac {1}{x-i\eta} - \frac {1}{x+i\eta}=2\pi i\delta (x)
\end{equation}
we obtain
\begin{equation}\label{lm2}
\langle S_{n}^z\rangle=\frac{1}{2}-
   \frac{1}{\Delta}
   \int
   \int dk_xdk_y
   \sum_{i=1}^{2N_z}\frac{D_{2n-1}(\omega_i)}
   {\mbox{e}^{\beta \omega_i}-1}
\end{equation}
where $n=1,...,N_z$.
As $<S_{n}^z>$ depends on the magnetizations of the neighboring layers via $\omega_i (i=1,...,2N_z)$,
we should solve by iteration the equations
(\ref{lm2}) written for all layers, namely for  $n=1,...,N_z$, to obtain the magnetizations of layers 1, 2, 3, ..., $N_z$
at a given temperature $T$. Note that by symmetry, $<S_1^z>=<S_{N_z}^z>$, $<S_{2}^z>=<S_{N_z-1}^z>$, $<S_3^z>=<S_{N_z-2}^z>$, and so on.
Thus, only $N_z/2$ self-consistent layer magnetizations are to be calculated.

The value of the spin in the layer $n$ at $T=0$ is calculated by

\begin{equation}\label{surf38}
\langle S_{n}^z\rangle(T=0)=\frac{1}{2}+
   \frac{1}{\Delta} \int \int dk_xdk_y
   \sum_{i=1}^{N_z}D_{2n-1}(\omega_i)
\end{equation}
where the sum is performed over $N_z$ negative values of  $\omega_i$ (for positive values the Bose-Einstein factor is equal to 0 at $T=0$).

The transition temperature $T_c$ can be calculated in a self-consistent manner by iteration, letting all  $<S_{n}^z>$  tend to zero, namely $\omega_i\rightarrow 0$. Expanding $\mbox{e}^{\beta \omega_i}-1\rightarrow  \beta_c \omega_i$ on the right-hand side of Eq. (\ref{lm2}) where $\beta_c=(k_BT_c)^{-1}$, we have by putting $\langle S_{n}^z\rangle=0$ on the left-hand side,
\begin{equation}\label{tcc}
\beta_c=2\frac{1}{\Delta}\int \int dk_xdk_y
   \sum_{i=1}^{2N_z}\frac{D_{2n-1}(\omega_i)}{\omega_i}
\end{equation}
There are $N_z$ such equations using Eq. (\ref{lm2}) with $n=1,...,N_z$.  Since the layer magnetizations tend to zero at the transition temperature from different values, it is obvious that we have to look for a convergence of the solutions of the equations Eq. (\ref{tcc}) to a single value of $T_c$. The method to do this will be shown below.

Let us show the numerical results of the above equations.

\subsubsection{Spin-wave spectrum}
\begin{figure}[htb]
\centering
\includegraphics[width=6.5cm,angle=0]{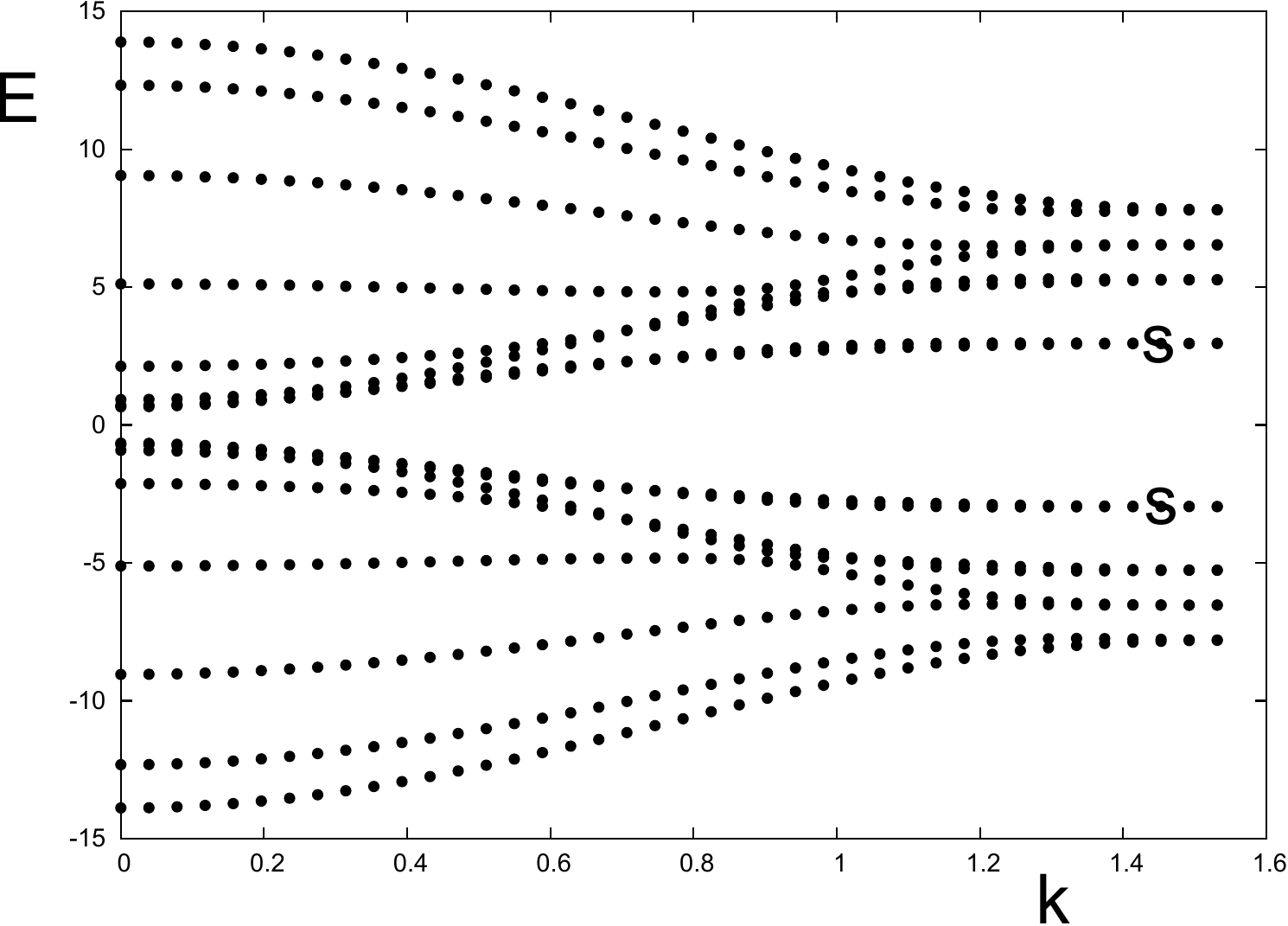}  % .eps
\includegraphics[width=6.5cm,angle=0]{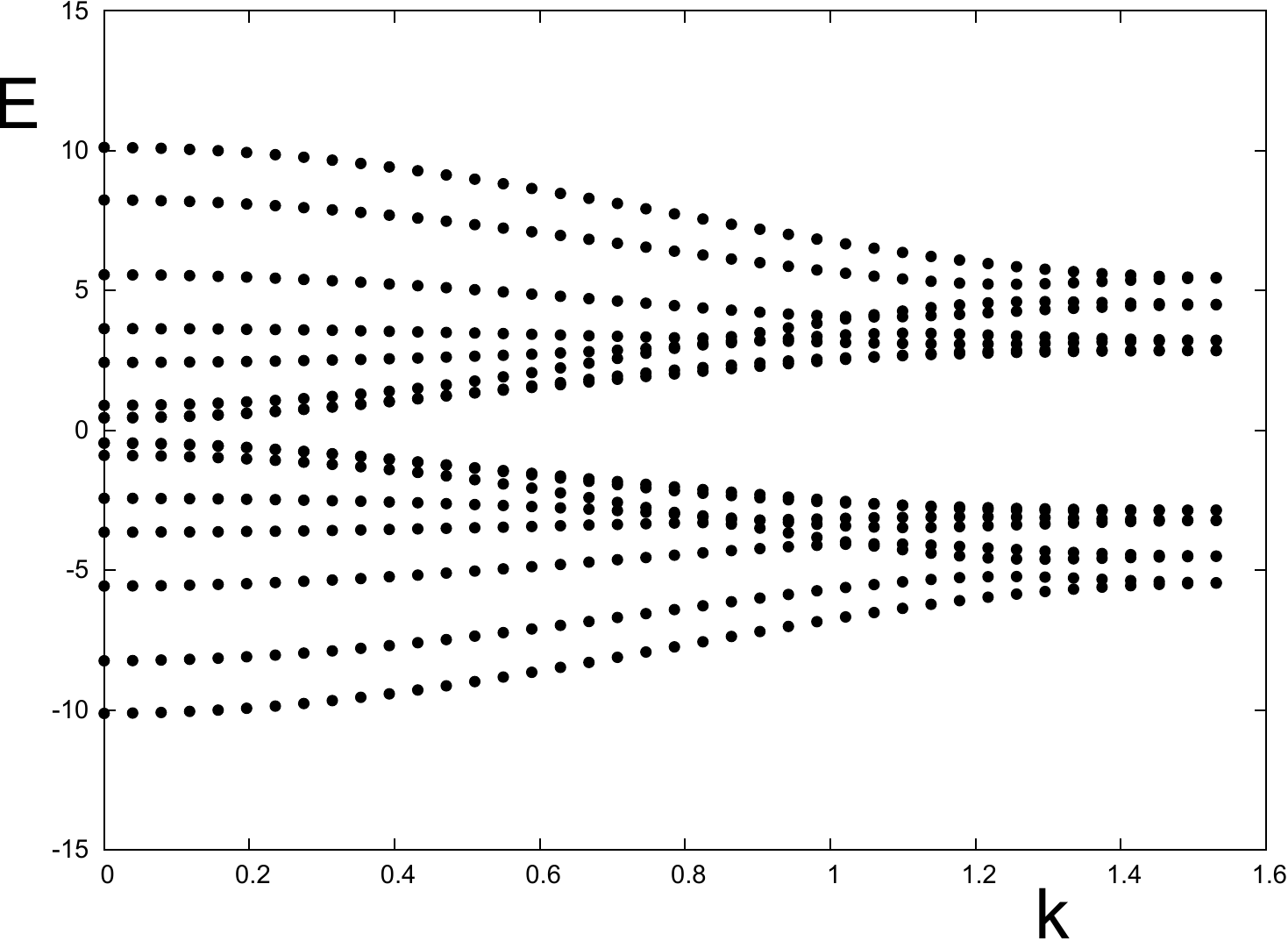}  % .eps
\caption{ Spectrum $E=\hbar \omega$ versus $k\equiv k_x=k_y$ for $J_2/J_1=-1.4$ at $T=0.1$ (top) and $T=1.02$ (bottom) for $N_z=8$ and $d=0.1$. The surface branches are indicated by $s$.}\label{sweta}
\end{figure}

Figure \ref{sweta} shows an example of the spin-wave spectrum $\omega$ obtained by solving det$|\mathbf M|=0$, as said above. The surface  spin waves are indicated by "S" on the figure. Surface spin waves are localized on the surface and damped when going to interior layers (see explanation in \cite{Diep1979}).
%Fig16

\subsubsection{Spin contraction at $T=0$}
\begin{figure}[htb]
\centering
\includegraphics[width=7cm,angle=0]{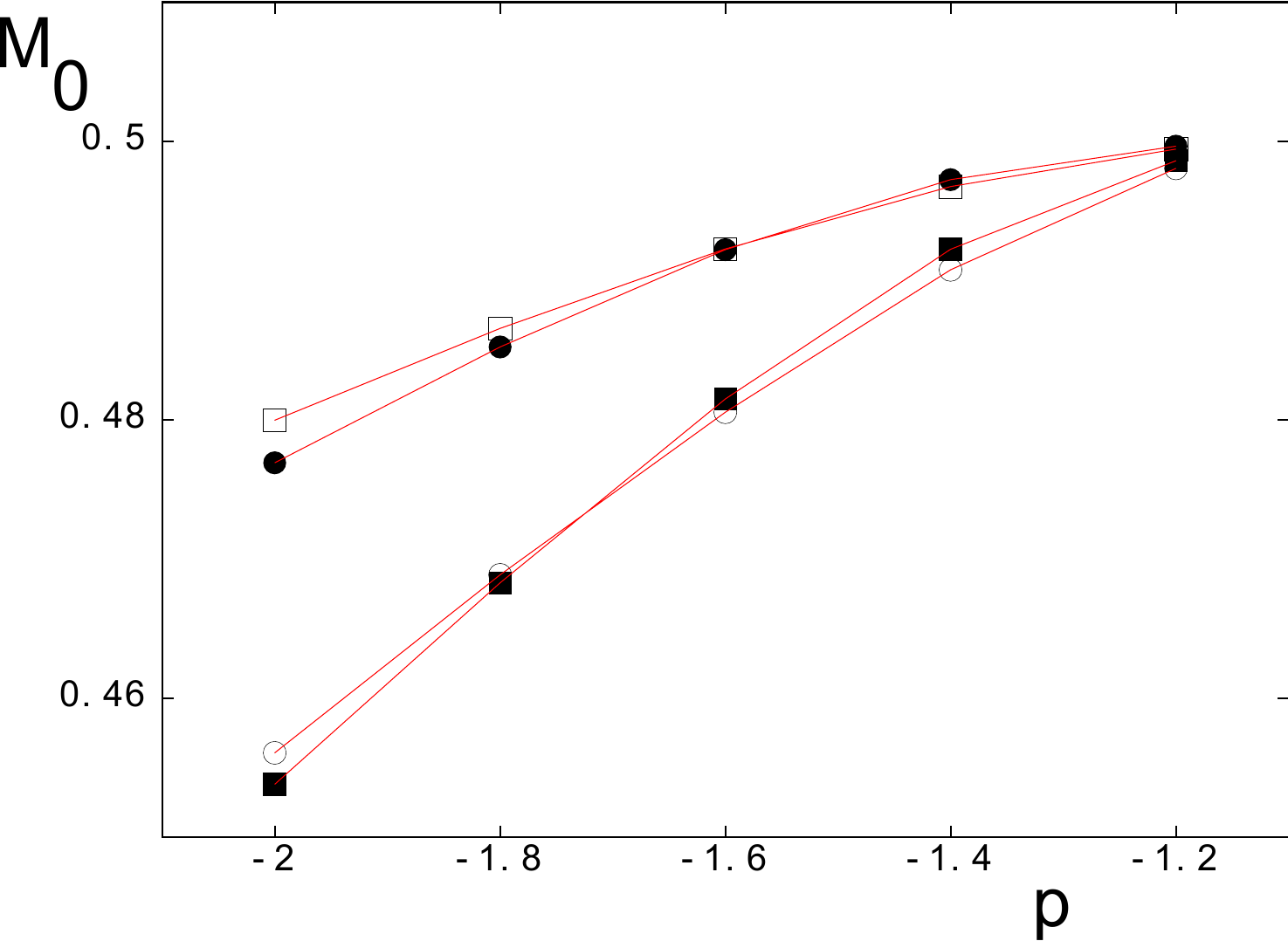}  % .eps
\caption{(Color online) Spin lengths of the first four layers at $T=0$ for several values of $p=J_2/J_1$ with $d=0.1$, $N_z=8$.
Black circles,  void circles, black squares and void squares are for first, second, third and fourth layers, respectively. See text for comments. }\label{spin0}
\end{figure}

It is known that in antiferromagnets, quantum fluctuations give rise to a contraction of the spin length at zero temperature \cite{DiepTM}.  We will see here that a spin under a stronger antiferromagnetic interaction has a stronger zero-point spin contraction. The spins near the surface serve for such a test. In the case of the film considered above, spins in the first and in the second layers have only one antiferromagnetic NNN while interior spins have two NNN, so the contraction at a given $J_2/J_1$ is expected to be stronger for interior spins. This is verified with the results shown in Fig. \ref{spin0}.   When $|J_2|/J_1$ increases, namely the antiferromagnetic interaction becomes stronger, we observe  stronger contractions. Note that the contraction tends to zero when the spin configuration becomes ferromagnetic, namely $J_2/J_1$ tends to -1.
%Fig17

\subsubsection{Layer magnetizations and transition temperature}

\begin{figure}[htb]
\centering
\includegraphics[width=7cm,angle=0]{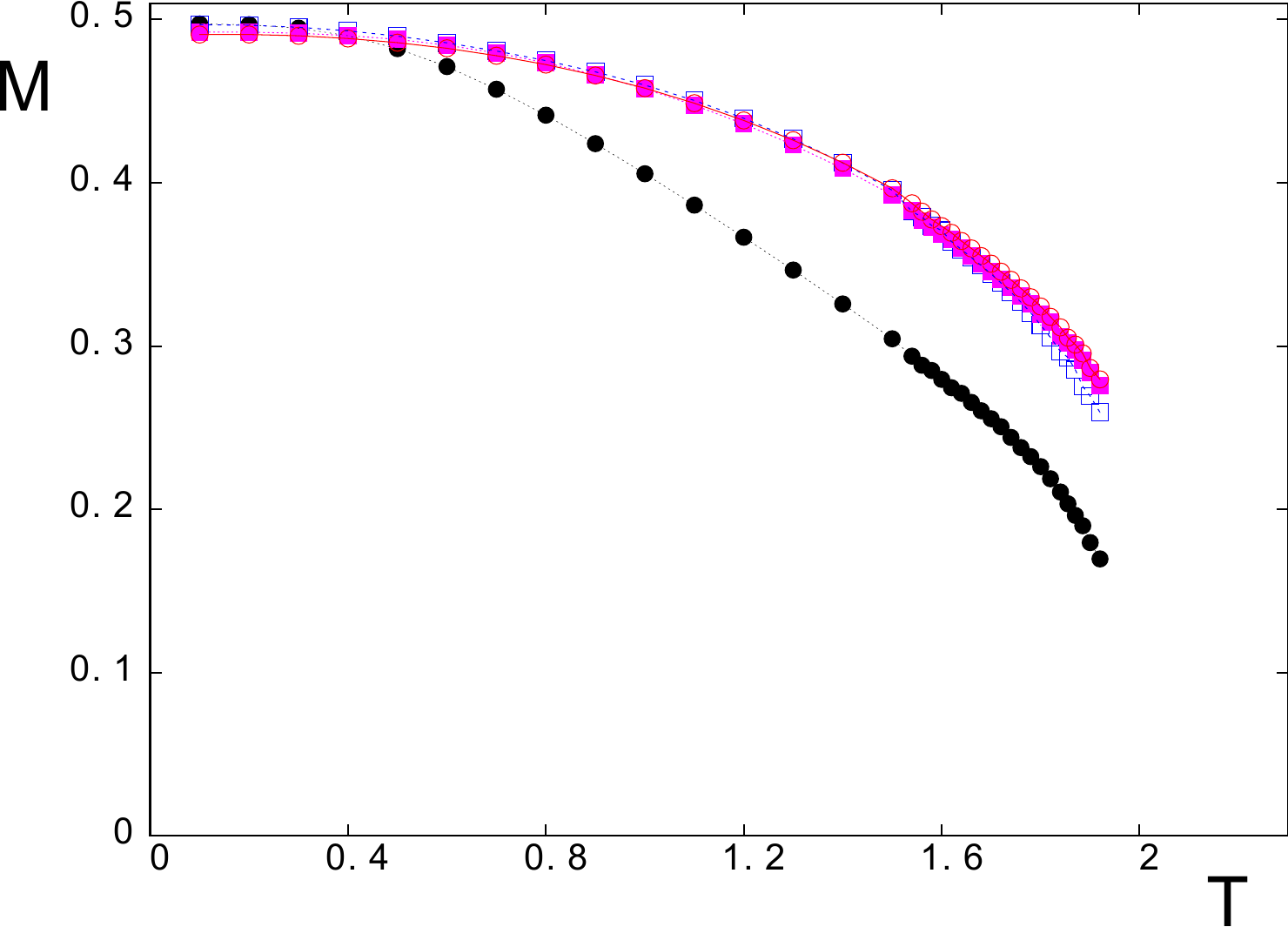}  % .eps
\includegraphics[width=7cm,angle=0]{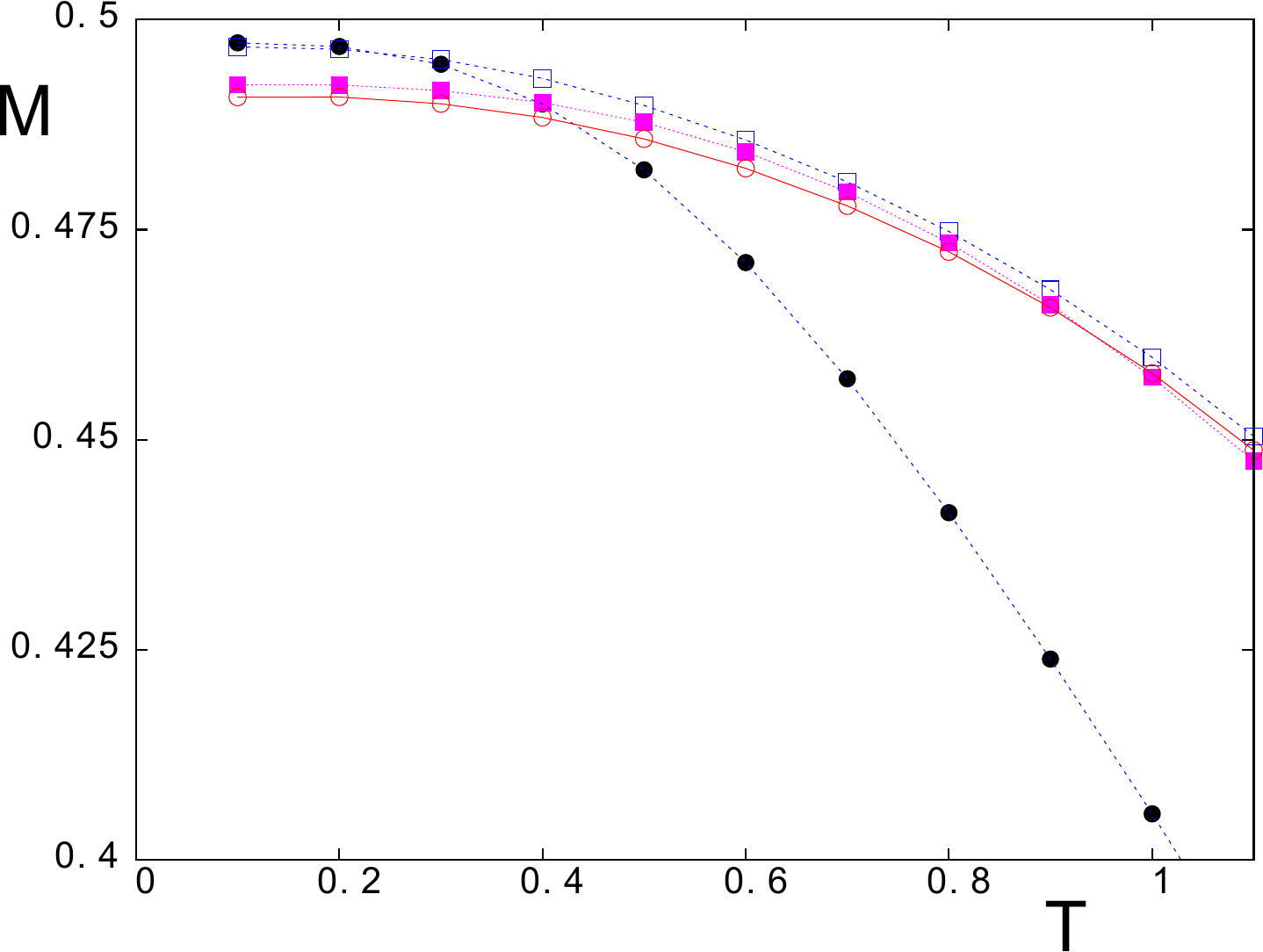}  % .eps
\caption{(Color online) Layer magnetizations as functions of $T$ for $J_2/J_1=-1.4$ with $d=0.1$, $N_z=8$ (top). Zoom of the region at low $T$ to show crossover (bottom). Black circles, blue void squares, magenta squares and red void circles are for first, second, third and fourth layers, respectively.  See text.}\label{magnet14}
\end{figure}

Let us show two examples of the magnetization, layer by layer, from the film surface in Figs. \ref{magnet14}  for the case where  $J_2/J_1=-1.4$ in a $N_z=8$ film.   Let us comment on these results:

(i) the shown result is obtained with a convergence of $1\%$. For temperatures closer to the transition temperature $T_c$, we have to lower the precision to a few percents which reduces the clarity because of their close values (not shown).

(ii) the surface magnetization, which has a large value at $T=0$ as seen in Fig. \ref{spin0}, crosses the interior layer magnetizations at $T\simeq 0.42$ to become much smaller than interior magnetizations at higher temperatures.  This crossover phenomenon is due to the competition between quantum fluctuations, which dominate low-$T$ behavior, and the low-lying surface spin-wave modes which strongly diminish the surface magnetization at higher $T$.  Note that the second-layer magnetization makes also a crossover at $T\simeq 1.3$.  Similar crossovers have been observed in quantum antiferromagnetic films \cite{DiepTF91} and quantum superlattices \cite{DiepSL89}.

Note that though the layer magnetizations are different at low temperatures, they will tend to zero at a unique transition temperature as seen below.  The reason is that as long as an interior layer magnetization is not zero, it will act on the surface spins as an external field, preventing them to become zero.  We have calculated self-consistently the transition temperature for each value of $J_2/J_1$ (see \cite{DiepHeli2015}).

%Fig18

Results of other values of  $J_2/J_1$ have been shown  in \cite{DiepHeli2015}. Effects of $d$, film thickness and surface exchange interactions different from the bulk ones have been shown in that reference.

\subsection{Other Applications in Frustrated Systems with Dzyaloshinskii-Moriya Interaction}
As seen in sections \ref{Frustration} and \ref{Skyr}, non-collinear spin configurations are found in many frustrated spin systems. The Green's function technique presented above can be applied to study elementary excitations and their thermodynamic effects.   We have seen above that although the formalism is rather cumbersome, the results cannot be obtained otherwise nowadays. 

We have applied the above Green's function technique to study for example several systems with Dzyaloshinskii-Moriya interaction in monolayer and bilayer magneto-ferroelectric materials \cite{Sharafullin2019} and in antiferromagnetic triangular lattice \cite{ElHog2022}.  The reader is referred to these papers for detailed results.

\section{Conclusion}\label{Concl}
I have reviewed in this paper a number of personal works on the frustration effects inspired  by a collaboration and numerous discussions with G\'erard Toulouse, at the beginning of my career, in the early 80's. I moved in the early 90's to the new Unversity of Cergy-Pontoise. I saw him again in the years 2000 when he was a member of the Orientation Council of our university. Needless to say, I was very happy to discuss again with him on physics and on other subjects of life as well.  

The search for frustration effects was uninterrupted in my activities to this day. As seen in this review, frustration is at the origin of many striking phenomena, going from the coexistence of order  and disorder in systems at equilibrium to the existence of reentrant phases and disorder lines, passing by systems with non-collinear spin structures and skyrmions. The frustration in materials opens many new areas of  physics not only on the theoretical point of view but also on many applications in spintronics, as mentioned above. This paper is just a personal account of the emergence of a modern physics in which various frustrated systems are still actively studied.  

To conclude, I would like  to emphasize that the concept of frustration is found also in other disciplines such as biology \cite{Millane} where  geometric frustration can drive the development of structure in complex biological systems, and may have implications for the nature of the actin-myosin interactions involved in muscle contraction.  In chemistry, the so-called "chemical frustration" allows to explain chemical properties of many complex compounds \cite{Frederickson,Harris}.  Even in  sociophysics where individuals are represented by spins, the frustration due to conflicts plays an important role in the dynamics of social groups \cite{Naumis,Kaufman19}.

%\acknowledgments
\vspace{2cm}

The author is grateful to his many collaborators and former doctorate students for contributions to works mentioned in this paper.

\end{document}